\documentclass[10pt,conference]{IEEEtran}
\usepackage{float}
\usepackage{hyperref}
\usepackage{algorithm}
\usepackage{algpseudocode}
\usepackage{graphicx}
\usepackage{textcomp}
\usepackage{xcolor}
\usepackage{soul}
\usepackage{textgreek}
\usepackage{caption}
\usepackage{tikz}
\usepackage{multirow}
\usepackage{textgreek}
\usepackage{wrapfig}
\usepackage{amsmath}
\usepackage{subfig}
\usepackage{enumitem,kantlipsum}

\hypersetup{
    colorlinks,
    citecolor=black,
    filecolor=black,
    linkcolor=black,
    urlcolor=black
}

\algdef{SE}[DOWHILE]{Do}{doWhile}{\algorithmicdo}[1]{\algorithmicwhile\ #1}%

\usepackage{xspace}
\newcommand{\artist}{{ArtISt-sim}\xspace}
\newcommand{\astra}{{ASTRA-sim}\xspace}

\begin{document}

\title{Dally: A network-placement sensitive cluster scheduler for  deep learning}




\author{\IEEEauthorblockN{Aakash Sharma, Vivek M. Bhasi, Sonali Singh, Mahmut T. Kandemir, George Kesidis, Chita R. Das}
\IEEEauthorblockA{Computer Science and Engineering, The Pennsylvania State University} 
sharma.aakash258@yahoo.com, vivekmbhasi@gmail.com, sonali0503@yahoo.co.in,
\{mtk2, gik2, das\}@psu.edu
\vspace{-4mm}}
\maketitle
\footnotetext{
This paper has been \textbf{accepted for presentation at the IEEE Big Data 2025 conference}.
}
\thispagestyle{plain}
\pagestyle{plain}

\begin{abstract}
Deep learning (DL) constitutes a significant workload within public or private cloud platforms. Organizations commonly engage in training multiple Deep Neural Networks (DNNs) in multi-tenant GPU clusters. The primary consideration in training such models is the associated ``cost'', which correlates directly with the GPU usage duration in the cluster. 
Previous studies have demonstrated that communication latency of interconnected GPU clusters could be a significant component of the overall training time and thus, minimizing the communication latency with intelligent scheduling of training jobs on physically close GPUs should significantly reduce the long training times. However, state-of-the-art cluster schedulers are rather agnostic to the ``proximity-based'' job consolidation. To compound the issue, different DNN models display varied communication latency even with similar levels of consolidation. This, in turn, results in prolonged training times.\par

Motivated by the observations above, we propose a novel GPU-cluster scheduler, Dally, for Distributed DL (DDL) workloads, that enables proximity-based consolidation of GPU resources based on the DDL jobs'  sensitivities to the anticipated communication-network delays. Dally consists  of three major components: 
(i) a mechanism based on the classical delay scheduling algorithm to facilitate job placement and consolidation; 
(ii) a network-sensitive job preemption strategy; and (iii)  an ``auto-tuner'' to optimize delay timers for effective delay scheduling. Additionally, to enable a cost-effective methodology for large-scale experiments, we develop a data-driven DDL cluster simulation platform, \artist.
Employing \artist, we compare Dally against several state-of-the-art alternatives on  real-world workload traces to demonstrate the benefits of our design. Dally can provide improvement in makespan of up to 69\% (68\% mean) across all training jobs, compared to the prevailing consolidation-based scheduling methods. The resultant boost in system throughput enables the reduction of average job completion time by up to 36\% (26\% mean) and decreases average communication latency by up to 83\% (66\% mean), under congested network conditions, as compared to  state-of-the-art.
\end{abstract}



\section{INTRODUCTION} \label{sections: introduction} 
Deep learning (DL) with large-scale cloud computing resources has facilitated the development of deep (very large) neural network (DNN) models, driving cutting-edge advancements in AI.  These models, powering breakthroughs in domains such as autonomous vehicles~\cite{self-driving}, multimedia processing~\cite{multimedia-DL}, generative AI~\cite{GAI}, neuromorphic computing \cite{nebula, skipper, gesture}, text classification \cite{Fast_Sentence_Classification} etc. often require training hundreds of millions, if not billions or trillions, of parameters to achieve the desired accuracy. To expedite model convergence within a reasonable time-frame, parallel training paradigms based on data parallelism~\cite{pytorch_dist} and model parallelism~\cite{megatron} are employed, simultaneously utilizing multiple GPUs (or related ``neural processing'' accelerators like TPUs~\cite{tpu}) to train a single model. Furthermore, AI enterprises or groups may train hundreds or even thousands of such DNN models~\cite{philly_trace_analysis}. In certain cases, a Machine Learning as a Service (MLaaS)~\cite{mlaas} provider could aggregate a substantial number of DL jobs from its diverse clientele into a single extensive GPU cluster. Note that MLaaS could be used both for training models from scratch or fine-tuning already-trained models.
However, GPUs are costly, and the prolonged training times of these models on GPUs can easily lead to exorbitant training expenses.\par

To alleviate the financial burden of DL, DNN workloads are executed on ``shared'' GPU clusters, such as those offered by public cloud providers. In such cases, the cost is determined by the wall clock time spent (makespan) utilizing cloud VMs.

Recent studies~\cite{pipedream,alibaba_pai, stash} have shown that communication latency is a major contributor to prolonged training times (and thereby, increased costs) in Distributed DL (DDL). This stems from the fact that GPUs within a cluster are {\em interconnected}  through various types of networks. During DNN training,  collective communication algorithms such as all-reduce, scatter-gather, and others are employed to exchange tensors among GPU workers.  This communication latency can vary significantly depending on the underlying network type connecting the GPUs. Generally speaking, the more proximal (physically close) the GPUs of a job are to each other within the network, i.e.,  the more ``consolidated'' they are, the lower the communication latency.\par

In distributed deep learning (DDL) scheduling, job placement consolidation plays a pivotal role in minimizing communication latency associated with parameter synchronization. A DDL job can be (primarily) consolidated in three typical ways: (i) placing all worker GPUs on the same (physical) machine; (ii) locating all worker GPUs on the same rack interconnected via Infiniband~\cite{Infiniband} technology; or (iii) spreading the worker GPUs across the network through inter-rack connections. 


\begin{figure}
    \centering
    \includegraphics[width=1\linewidth, height=0.5\linewidth]{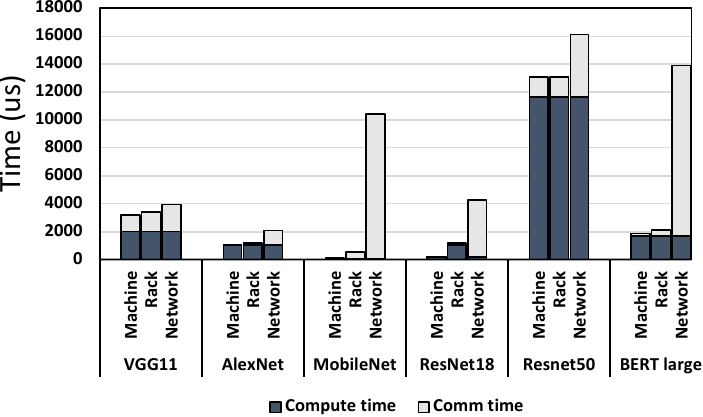}
    \caption{Single iteration training time for models consolidated on the same machine, rack, and across the network. Latency increases as the GPU workers grow physically apart.}
    \label{fig:itr_run_time}
\end{figure}

Each of these ``network tiers'' introduces significantly different communication latency, with the least latency being on the same machine, followed by the rack, and then the network.
To illustrate this, we run a single training iteration of some popular DNNs using 8 GPUs placed across the same -- (i) machine, (ii) rack and (iii) across the network, and measure their running times (shown in Fig.~\ref{fig:itr_run_time}).
We observe that as the GPU workers become less physically proximal on account of the underlying network placement configuration, the training latency increases.
Furthermore, this increase in training latency can be attributed to the higher communication overhead incurred by the respective network.
Section~{\ref{sections: motivation_job_consol}} provides further details on the experimental setup.
To mitigate the communication latency, a straightforward approach involves consolidating all GPUs of a job as much as possible. However, due to multi-tenancy, when cluster managers attempt to consolidate a job on physically proximal GPUs, they often encounter delays while waiting for resources to become available, leading to increased queueing times. 
Furthermore, previous research~\cite{tiresias, stash} has clearly demonstrated that DNN models exhibit varying degrees of ``network sensitivity,'' implying that each DNN model would experience a different amount of increase in communication latency as network conditions deteriorate. Thus, not all models would equally benefit from communication latency reduction through consolidation, but they would all incur the same queueing delay. Hence,  it is crucial to consolidate DNN jobs based on their ``network sensitivities'' i.e., relative communication latency, in order to {\em minimize}  queueing delays arising from consolidation.\par

A prior cluster scheduler, Tiresias~\cite{tiresias}, implements a consolidation strategy for DDL jobs based on the ``skew'' of the model, which is defined as the ratio of the largest tensor size to the overall model size. In this strategy, jobs with high skew are consolidated into as few machines as possible, while others are left unconsolidated. Another prior approach~\cite{sc17-topoaware} either ``packs'' or ``spreads'' training jobs across the data-center. 
These approaches may lack flexibility for ``moderate consolidation,'' such as placing a job on the rack for a ``moderately network-tier''--sensitive model. Moderate consolidation is now more viable for communication due to the availability of high speed contemporary networking hardware. 

The contemporary Infiniband-based~\cite{Infiniband} rack switch offers remarkably high bandwidth, reaching up to 400 Gb/s per port with the utilization of GPU RDMA~\cite{gpudirect-RDMA}, effectively reducing the rack-level communication latency. Moreover, even modern network (Ethernet-based) switches provide 800 Gb/s bandwidth per port, albeit with higher link-latency ~\cite{nvidia-spectrum}. Yet, DDL cluster schedulers remain mostly oblivious to such advancements and are unable to harness the consequent potential performance improvements. To address these limitations, schedulers should be equipped with the ability to dynamically adjust consolidation based on the network sensitivity of individual jobs. This adjustment should take into account the diverse performance characteristics of different network tiers, leveraging the capabilities of modern networking hardware essential for lowering communication latency.



To achieve these goals, we propose {\bf Dally}, a holistic cluster scheduler for DL job placement that accounts for the underlying network topology, as well as the network sensitivity of each job to make optimal placement decisions. Dally accomplishes this by employing the classical ``delay scheduling''    algorithm~\cite{delay_scheduling} in the context of DDL workloads, coupled with a novel network-sensitive preemption strategy for job placement and consolidation.  
Specifically, for every DDL job, Dally consults its dedicated ``delay timers'' to decide whether to accept a resource offer at the offered network-tier {\em or} wait in anticipation of an offer with better network-tier until a delay timer has elapsed. Traditionally, the delay timers are hand-tuned and remain fixed for a given system.
However, we introduce a novel ``auto-tuner'' that dynamically evolves with the system to tune these delay timers for high-efficacy delay scheduling. With the help  of these techniques, Dally is able to outperform state-of-the-art schedulers in terms of makespan, job completion time (JCT), and communication latency.

Additionally, for an accurate evaluation of Dally, we have developed a simulation platform -- \artist.
The use of simulation is motivated by the fact that development of novel DL cluster schedulers is currently impeded by the exorbitant costs associated with the large-scale experimental setups needed.  A single GPU for DL  costs thousands of dollars, and replicating a real-world DL data center necessitates a substantial number of these GPUs for extended periods. \artist has been developed by extending a prior work~\cite{themis-nsdi} and utilizing \astra~\cite{rashidi2020astra}, a highly accurate DDL job simulator. Our simulation platform possesses two key capabilities: (i) it can dynamically determine network slowdowns based on the specific placement of a DDL job (its assigned GPUs) within the simulated cluster, and (ii) it can simulate a datacenter equipped with modern networking hardware (e.g., NVIDIA Quantum~\cite{nvidia-quantum} and Spectrum switches~\cite{nvidia-spectrum}),  which is crucial for the purposes of this work. This combination of features enables \artist to provide a realistic and accurate representation of modern DDL workloads.  Moreover, to the best of our knowledge, modern high speed network switches such as the ones stated before are unavailable in any public cloud for use as a test-bench.  Hence, \artist can enable researchers  to develop novel DDL scheduling techniques with modern networking hardware, facilitating effective research in DDL cluster scheduling. \par




To the best of our knowledge, Dally is the first job scheduler, suitable for public cloud deployment, which introduces network-sensitive GPU consolidation policy for current and next-gen fast network hardware. This is achieved through a combination of a novel network-sensitive preemption policy and an adapted delay scheduler with an auto-tuning mechanism. Past DL cluster schedulers suffer from:  (i) lack of awareness of various network tiers which prevents them from utilizing the  medium-bandwidth connectivity options, such as GPU RDMA; (ii) rigid consolidation policy restricting job placement to either occupying the fewest possible machines or dispersing them across the network; (iii)  inaccurate or no information about the job's sensitivity to network stalls and; (iv) sub-optimal job prioritizing. This work also introduces a novel simulation platform for high-fidelity DL cluster simulations, \artist.

To summarise, in this paper, our \textbf{primary contributions} can be outlined as follows:

$\bullet$ We present Dally, a novel DL cluster scheduler, crafted through a co-design of networking hardware and underlying software, to reduce end-to-end makespan, average job completion time (JCT), and communication latency. Dally employs delay scheduling to achieve optimal job consolidation and placement and implements a network-sensitive preemption policy, facilitating prioritized placements of jobs with heightened sensitivity to network conditions.

$\bullet$ Dally also incorporates an ``auto-tuner'' capable of {\em dynamically adjusting} the delay timers, based on network usage, required for delay scheduling.

$\bullet$ Further, we develop a high-fidelity (iteration-level) DDL cluster simulator using \astra called \artist \cite{artist-sim}, which effectively simulates a DDL datacenter equipped with contemporary networking hardware.

$\bullet$ Finally, extensive evaluation of Dally using \artist, encompassing both batched and Poisson
workloads of six real-world DNN models, reveals significant improvements. Specifically, Dally improves makespan by over 69\% (68\% mean) compared to state-of-the-art consolidation based policies, reduces average JCT by more than 36\% (26\% mean), and minimizes communication latency by up to 83\% (66\% mean), under congested networking conditions.

The rest of this paper is organized as follows. 
In Section~\ref{sections: background}, we discuss the background pertinent to delay scheduling and our simulation. In Section~\ref{sections: motivation}, we motivate our problem and our scheduler scheme is described in Section~\ref{sections: design}.
The experiment methodology and evaluation is presented in Sections~\ref{sections: methodology} and~\ref{sections: evaluation}.
Related work is in Section~\ref{sections: related-work} and finally,  Section~\ref{sections: summary} summarizes our major observations and findings.

\section{BACKGROUND} \label{sections: background} 


\subsection{Delay Scheduling Based on Data Locality}
Delay scheduling is a traditional scheduling algorithm that defers the scheduling of a job to the future when a suitable/preferred job placement destination (hardware) is not immediately available.  Instead of promptly accepting a resource offer from any worker machine, a job undergoes a ``delay time'', anticipating the availability of a worker node with enhanced data locality in the future. Data locality refers to the location of the job's input data with respect to its compute hardware, which could be on the same machine, the same rack, or somewhere accessible over the network.
System administrators set the delay time, often maintaining the default values established by cluster schedulers. Traditional schedulers, like YARN~\cite{yarn} and Spark~\cite{spark}, typically implement delay scheduling as a ``default'' strategy for better data locality.

\subsection{Simulation Platforms} 
\astra (Accelerator Scaling for TRAining Simulator) is a comprehensive simulation platform facilitating parameterized descriptions of DNN model, system, and network fabric for end-to-end simulation of a DDL loop. 
\astra users can configure various DDL training setups, including DNN models (number and types of layers, GEneral Matrix Multiplication (GEMM) size \cite{gemm}, etc), DDL parallelism (data, model, hybrid, etc.), fabric design (number of links, latency/bandwidth per link, etc.), and fabric topology (2D/3D torus, switch, ring, etc).  \astra leverages existing network-layer simulators, e.g., Garnet \cite{garnet}, Analytical \cite{analytical}, and ns-3 \cite{rashidi2020scalable}, for packet-accurate simulations. An improved version of \astra \cite{won2023astrasim20} with capabilities such as arbitrary model parallelism, improved hierarchical topology parameterization and enhanced memory system modeling is now available. Also, \cite{Akella22} developed a simulator using \astra and ns-3~\cite{rashidi2020scalable} to model various congestion control schemes.

\astra expects the following three input files to simulate a single loop of DDL training:\\ 
$\bullet$ {\bf Workload file:} This file contains the delay in cycles for the GEMM operations of each layer during the forward and backward pass along with layer-wise communication size.\\
$\bullet$ {\bf System file:} 
The system file provides parameters to the network scheduler  such as topology, packet scheduling policy (e.g., LIFO,  FIFO), all-reduce implementation, etc.\\
$\bullet$ {\bf Network file} The network files encompass details regarding the physical network topology, network dimensions, link information (bandwidth, latency, etc.) etc.\par

We want to emphasize however that \astra functions as a single DDL job simulator and lacks the capability to simulate a large GPU cluster running multiple DL jobs.  
Hence, in order to accurately replicate the behavior of a real-world data center, a multi job deep learning simulator is essential.





\section{MOTIVATION} \label{sections: motivation} 
\begin{figure}
    \centering
    \includegraphics[width=1\linewidth]{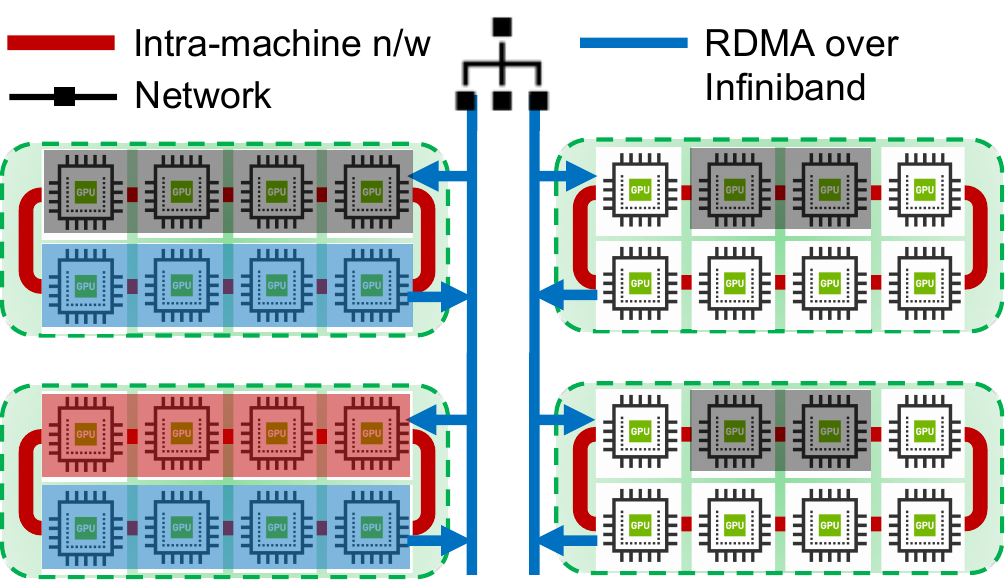}
    \caption{A typical (hierarchical) datacenter network (n/w).}
    \label{fig:nw_arch}
\end{figure}

A typical DL cluster today comprises a large number of GPUs communicating via different types of network interconnects. These interconnects may be (i) intra-machine like NVSwitch~\cite{nvlink-nvswitch}; (ii) intra-rack connected via Infiniband; or, (iii) inter-rack connected via modern network (Ethernet) switches.
Such a hierarchical network structure is depicted in Fig.~\ref{fig:nw_arch}. In this figure, a job mapped to GPUs on the same machine is shown in red; a job running on the same rack (but on different machines) is marked in blue; and a job mapped to GPUs across the network is shown in grey.
The communication latency is lowest for the jobs highlighted in red, followed by those in blue and then grey. 

\subsection{Communication Latency in DDL Clusters}
The effectiveness of DDL training throughput is heavily influenced by the essential role of communication speed and latency due to the significant volume of data transfer among DNN workers during gradient synchronization.
As highlighted in~\cite{kungfu}, an 8-GPU server engaged in training a ResNet50  model produces 4 GB of gradients per second. Also, previous studies such as~\cite{pipedream,stash} reveal that the communication latency can be as high as $5\times$ the actual compute time. 
Under a layer-wise model synchronization regime, starting from the output layers under back-propagation: (i) gradients (of the loss training-objective based on the GPUs' assigned training-data batches) are computed with respect to parameters of the current layer, (ii) these gradients are then exchanged among the GPUs, and finally, (iii) the current layer's parameters are simultaneously updated by each GPU.
Hence, multiple DDL models can quickly overwhelm the link bandwidth.
This underscores the importance of high-speed and low-latency interconnects between GPUs for mitigating communication-related stalls.



\subsection{Job Consolidation in DL Clusters}\label{sections: motivation_job_consol}
Cluster schedulers like Tiresias \cite{tiresias} aim to reduce communication latency by consolidating job placement, and assigning GPUs to a job within the fewest machines possible. However, due to multi-tenancy, such consolidation can lead to lengthy queueing delays, depending on the cluster's load. Individual jobs may reject resource offers generated by the cluster scheduler if the offered GPUs are not located on the same machine or rack, especially when  GPU demand is high.  As a result, jobs remain queued, increasing their job completion time (JCT). 

However, not all jobs benefit equally from consolidation, as network latency varies across DNN models. More specifically, jobs which are less sensitive to network quality (less `network-sensitive' models) can tolerate less stringent network consolidation (consolidation on physically distant GPUs) compared to highly ``network-sensitive'' jobs. 
To illustrate this, in Fig.~\ref{fig:itr_run_time}, we simulate the single iteration DDL running times using \astra for various models with different network placements. 
Each model is simulated using 8 GPUs wherein the GPUs are placed -- (i) within the same machine, (ii) within the same rack, and (iii) distributed across racks. As expected, in general, consolidation on physically proximal GPUs yields lower run times. However, we observe significant variations in communication latency  across machine, rack and network for different DNN models. 
Table~\ref{tab:overhead_table} illustrates the communication time (latency) expressed as a percentage of the compute (GPU) time. The significant variations in latency, such as ResNet-18 exhibiting a notable rack level latency, stem from the unique characteristics of individual models or their network sensitivity. Prior work~\cite{stash} has examined the features of DNN architectures that influence latency behavior in relation to the underlying network topology.  Thus, model network  sensitivities could act as a potentially relevant metric in determining the priority of each job with respect to consolidation: \emph{the higher the network sensitivity of a model, the more it should be prioritized to get an allocation with physically proximal (consolidated) GPUs.}\par

In this direction, Tiresias~\cite{tiresias}, seeks to categorize network-sensitive jobs by employing model ``skew,'' which is defined as the ratio of the largest tensor size to the overall model size. Tiresias  consolidates jobs with high skew, while other jobs accept any resource offer they receive. The underlying assumption is that skew accurately indicates network sensitivity. However, using \astra to analyze the communication latency of various models under different network placements and high speed networking hardware reveals {\em weak correlation} between skew and network sensitivity as shown in Table~\ref{tab:overhead_table}. For example, ResNet18 has low skew but very high network level communication latency and VGG11 has high skew but low communication latency. This could be due to the rapid advancements in state-of-the-art (SOTA) network hardware which has changed the various latency incurred during gradient exchange. 
A plausible relationship between the DNN model and communication latency is that the latter depends on the gradient size per layer and the underlying network characteristics such as bandwidth and latency \cite{stash}. 

\begin{table}
\small
\begin{tabular}{ccccc}
\hline
Model       & Skew & \begin{tabular}[c]{@{}c@{}}Machine\end{tabular} & \begin{tabular}[c]{@{}c@{}}Rack\end{tabular} & \begin{tabular}[c]{@{}c@{}}Network\end{tabular} \\ \hline
VGG11 \cite{vgg}       & High & 1\%                                                        & 6\%                                                     & 7\%                                                        \\
AlexNet \cite{alexnet}     & High & 2\%                                                        & 13\%                                                    & 100\%                                                      \\
MobileNetV3 \cite{mobilenet} & High & 42\%                                                       & 940\%                                                   & 19592\%                                                    \\
ResNet18 \cite{resnet}    & Low  & 7\%                                                        & 116\%                                                   & 2749\%                                                     \\
ResNet50 \cite{resnet}    & Low  & 12\%                                                       & 12\%                                                    & 38\%                                                       \\
BERT large \cite{bert}  & Low  & 8\%                                                        & 23\%                                                    & 715\%                                                      \\ \hline
\end{tabular}
\caption{Communication latency as a \% of compute time. }
\label{tab:overhead_table}
\end{table}

\subsection{Improvements in SOTA Network Hardware} 
Recent advancements in network hardware have introduced remarkably high network bandwidth with exceptionally low latency. Hardware manufacturers, such as NVIDIA, are now offering network switches that deliver several-fold improvements in QoS compared to previous-generation hardware. For intra-machine networking, NVSwitch~\cite{nvlink-nvswitch} provides a bandwidth of up to 900 Gb/s, surpassing PCIe's bandwidth of only up to tens of Gb/s combined with high latency \cite{li2019evaluating}. Furthermore, for  Infiniband-based intra-rack networking, NVIDIA Quantum switches now offer up to 400 Gb/s bandwidth per port with ultra-low latency using GPU direct RDMA. In contrast, standard Infiniband offers only up to 50 Gb/s bandwidth while experiencing relatively high latency (up to tens of microseconds)~\cite{Infiniband-eval}.  Finally, NVIDIA Spectrum Ethernet switches  provide up to 800 Gb/s bandwidth for inter-rack networking. Note that ethernet switches suffer from high latency. However, \emph{current schedulers lack awareness of such high-speed network hardware, which could be harnessed for improved performance.} This underscores the demand for a novel approach to DL cluster scheduling. 

\subsection{Limitations of Current SOTA DL Cluster Schedulers}
Current SOTA schedulers like Tiresias~\cite{tiresias} and Gandiva~\cite{gandiva} have several shortcomings such as: (i) \textit{leveraging only limited knowledge of the datacenter network}: due to the lack of awareness of the various network tiers, e.g., the medium-bandwidth GPU direct RDMA that exists within a rack, these schedulers often end up making a sub-optimal job placement decision, in this case, mapping the job across multiple racks;  (ii) \textit{a strict consolidation  policy}: restricting the job placement to either the fewest machines possible (for high skewed models) may lead to high queueing delays in case the former is not available. This necessitates an intermediate approach, allowing for a gradual and graceful ``relaxation'' of consolidation, particularly for moderately sensitive jobs based on queueing delay and network sensitivity. 
(iii) \textit{Sub-optimal job priority}: Tiresias employs a generalized version of Least Attained Service (LAS)~\cite{las} called ``Discretized 2D-LAS'' (2DAS)~\cite{tiresias} to determine job priority. 
The 2DAS value for a job is calculated by multiplying the time for which it has been running by the number of GPUs it utilizes. Tiresias prioritizes jobs based on their 2DAS values, with lower values indicating higher priority. Consequently, a job with the highest 2DAS value is least likely to be scheduled for execution and most likely to be preempted out. 
However, a job with unfavorable network placement and high network sensitivity may have been running for an extended period but may not have made significant progress due to the excessive network latency associated with its placement. This would lead to a high 2DAS value and make it less likely to receive resource offers that are consolidated, as jobs with lower 2DAS values would receive the most favorable offers.

These limitations of existing DL cluster schedulers underscore the need for a new approach that incorporates {\em both} network awareness and network sensitivity-based job consolidation. However,  development of such scheduler for cutting-edge network technology is hindered by the high cost of the associated experimental setups and the availability of such hardware, necessitating a high-fidelity cluster simulator. 

\subsection{A high-fidelity simulator for DDL scheduling research} 
The exorbitant cost of deep learning GPUs, with several GPUs required to create a large deep-learning cluster, poses a significant challenge in conducting real-world experiments. This, combined with the need to run such experiments for extended periods, can escalate the cost to the hundreds of thousands of dollars.  
Even utilizing  public cloud resources may prove uneconomical as deep learning machines like the AWS P4 instances incur hourly charges of  more than \$32/hour \cite{aws_pricing}. 
Further complicating matters, research based on cutting-edge network hardware (such as the NVIDIA Quantum and Spectrum switches) necessitates a simulation platform due to its limited availability.  
To the best of our knowledge, no public cloud provider offers access to such hardware.  
To address these challenges, research efforts like~\cite{rashidi2022themis} have turned to \astra as a viable tool for simulating the latest network hardware. However, \astra is a single DDL job simulator.\par

Existing multi-job DDL cluster scheduler simulators, such as~\cite{themis-nsdi}, suffer from low accuracy due to their unrealistic approach to network latency modeling. For example, \cite{themis-nsdi} implements static slowdown penalties of job placement on the rack and the network \emph{irrespective} of the DNN model or network hardware.
But, to achieve high confidence in the simulation results, a DL cluster scheduler simulator must {\em dynamically} determine the accurate network latency of a job based on specific placement within the cluster and the type of network hardware used to accurately reflect real-world conditions. In response  to this requirement, we propose a ``multi-job'' DL cluster simulator that employs \astra to accurately determine network latency.

\section{DALLY DESIGN} \label{sections: design} 
We now describe Dally, a novel network-placement sensitive DL cluster scheduler that leverages the principles of delay scheduling and a novel network-sensitive preemption policy. 
In the following sections, we delve into the details of our system design, including the simulation methodology employed to evaluate its effectiveness.

\subsection{Overview} 
\subsubsection{System goal}
The system processes a stream of DL/DDL jobs that need to be scheduled for execution. These jobs exhibit heterogeneity in terms of the DNN model, the number of iterations required to attain the target accuracy, the compute time per iteration, and the GPU demand per job. 
The system may encounter jobs arriving in batches so that the aggregate GPU demand across all jobs would normally surpass the total number of GPUs available in the cluster. 
Alternatively, jobs may arrive according to a Poisson distribution.
Our goal is to construct a DL cluster scheduler that aligns with the target objectives and assumptions elaborated upon in the subsequent sections. 
This scheduler will function within a cluster comprising homogeneous GPUs connected via latest-generation network hardware in a hierarchical structure, as discussed in Section~\ref{sections: motivation}.
Generally, the scheduler is intended for an MLaaS provider who may have little to no knowledge about the DNN models being trained
including any historical data associated with it.

\subsubsection{Objective} 
The overarching goal of our scheduler is to minimize cloud expenditure for its users by reducing the time spent occupying cluster GPUs. 
This objective is achieved through two primary strategies.
\noindent
\begin{enumerate}[wide, labelwidth=!, labelindent=0pt]
\item Minimizing the end-to-end makespan for training multiple DNN models when submitted as a batch. 
\item Reducing job completion time (JCT) for individual jobs.
\end{enumerate}

\subsubsection{Constraints}
Dally must achieve the aforementioned objectives subject to the following constraints: 
\begin{enumerate}
\item \emph{Unknown JCT distributions}: We assume that the prior running time of any job is unknown.
\item \emph{Absence of model architecture information}: The scheduler lacks the knowledge of the specific DNN model architecture associated with each job.
\item \emph{Unknown job submission pattern}: The scheduler lacks information about the order and frequency of job arrivals. 
\item \emph{Unknown GPU demand}: The GPU resource requirements for individual jobs and the overall GPU demand across all running jobs are unknown beforehand. 
\end{enumerate}


\subsection{Design of the Dally Scheduler} 
\begin{figure}
    \centering
    \includegraphics[width=0.8\linewidth]{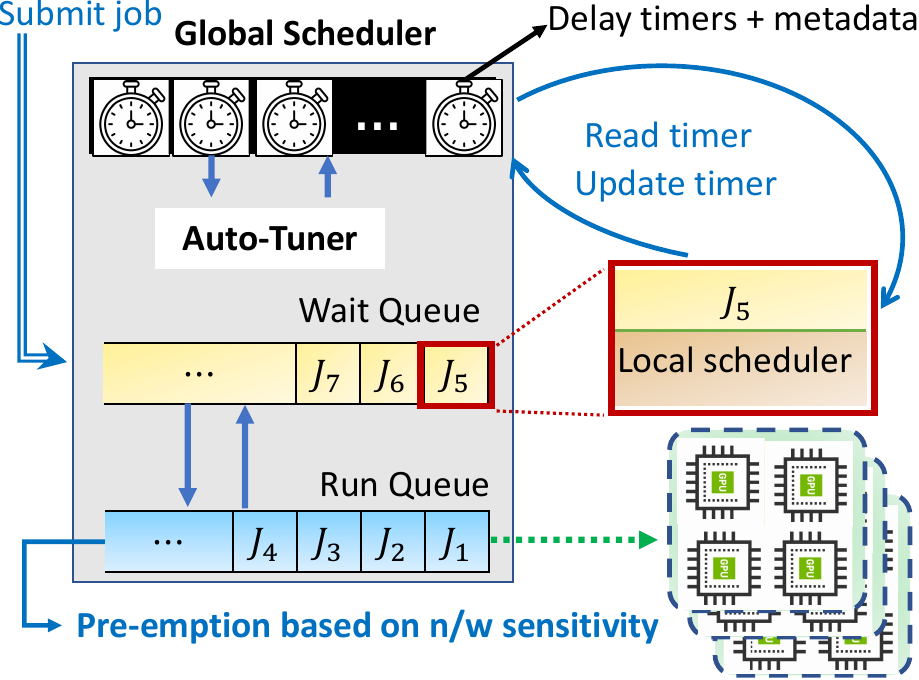}
    \caption{Scheduling scheme.}
    \label{fig:scheme}
\end{figure}

As stated earlier, Dally is designed to minimize cloud expenditure. To achieve this objective, it uses delay scheduling for job placement and employs a novel preemption strategy that incorporates network sensitivity. When a job is submitted to the cluster, the global scheduler (refer to Fig.~\ref{fig:scheme}) initially places it into a wait queue. The  scheduler then periodically generates ``resource offers'' to jobs in the wait queue based on the availability of cluster resources. The job's local scheduler (Fig.~\ref{fig:scheme}) makes a decision to accept or reject the offer based on delay timers which is maintained for each network dimension. Upon acceptance, the job is moved to the run queue, where it executes until it is either preempted or completes its execution.  Upon preemption, the job is {\em reintroduced} into the wait queue (after saving its current state) and remains there until it accepts another resource offer. The job then executes from its last saved state which consists of the model parameters, the state of the optimizer (e.g., current step size) and the number of iterations completed. Upon restart, the model state is first restored from the persistent storage and the job executes from the last iteration completed. Note that the  restore time is negligible compared to the long running times of DL jobs. An instance of Dally's mechanisms governing preemption and job placement is elaborated upon in the following sections. 
                              
\subsubsection{Preemption Priority} \label{sections: preemption} 
As already discussed in Section~\ref{sections: motivation}, Tiresias's  2DAS based priority mechanism is oblivious to network slowdowns. In a large networked cluster, an effective priority mechanism should consider the relative network sensitivity of jobs and the slowdowns they have experienced. Therefore, we propose a network-sensitive priority metric to address this shortcoming. We define the number of iterations completed as $I_{compl}$ and the total expected iterations as $I_{total-expected}$. Our approach necessitates the awareness of the overall expected iteration count which in general is a user input (hyper-parameter). Alternatively, users  can estimate the total iteration count needed for convergence through established extrapolation techniques such as~\cite{optimus}. 
It is important to emphasize that only an approximate value for the total expected iterations suffices.

Next, we define $T_{run}$ as the time spent by the job in the run queue and the total ideal job running time $T_{total-ideal-run}$ as the compute time per iteration of a job multiplied by the total number of expected iterations. The compute time per iteration can be measured by running a single training iteration of the model on a single GPU, where the iteration comprises the forward pass and backward pass (we assume no communication latency for ideal JCT calculations).
Profilers like DS-analyzer~\cite{ds-analyzer} may be used to find the exact compute time of an iteration. We now define work completed $W_{compl}$ and normalized running time $T_{norm}$ as:
\begin{align*}
W_{compl} = \frac{I_{compl}}{I_{total-expected}} \qquad
T_{norm} = \frac{T_{run}}{T_{total-ideal-run}}. 
\end{align*}

\noindent
Finally, we define our network sensitive priority $Nw_{sens}$ as:
\begin{align*}
Nw_{sens} &= \frac{W_{compl}}{T_{norm}}.
\end{align*}

A lower value of $Nw_{sens}$ indicates that the job has endured greater network-induced slowdowns, and, consequently, its priority should be increased. 
Resource offers are hence made in increasing order of $Nw_{sens}$.  This allows jobs with lower $Nw_{sens}$ to have a higher probability of receiving more consolidated resource offers. 

\subsubsection{Job Placement} 
\begin{algorithm}
\caption{1 On Resource Offer} 
\begin{algorithmic}[1]
\Require Resource offer $R$
\Ensure Accept or reject offer


\State $T_{last\_assignment}$ = $\mbox{get\_last\_resource\_assignment\_time}()$
\State $T_{starvation} = T_{curr\_time} - T_{last\_assigment}$
\State $g_d = \mbox{GPUs demanded by the job}$
\State $T_{Mc}$, $T_{Rk}$ = $\mbox{get\_tuned\_timers}(g_d)$


\If{$g_d \leq \mbox{GPUs available on a single machine in } R$}
\State Allocate $g_d$ GPUs on machine
\State $\mbox{update\_demand\_delay}(machine, t_{starvation}, g_d)$
\State $\Return$ \textbf{accept resource}
\EndIf

\If{$T_{starvation} < T_{Mc}$}
\State $\Return$ \textbf{reject offer}
\EndIf

\If{$g_d \leq \mbox{GPUs available on a single rack in } R$}
\State Allocate $g_d$ GPUs on rack
\State $\mbox{update\_demand\_delay}(rack, t_{starvation}, g_d)$
\State $\Return$ \textbf{accept resource}
\EndIf

\If{$T_{starvation} < T_{Rk}$}
\State $\Return$ \textbf{reject offer}
\EndIf

\State Allocate $g_d$ GPUs on network
\State $\Return$ \textbf{accept offer}
\end{algorithmic}
\label{algo: on_resource_offer}
\end{algorithm}

In a DL cluster, jobs may receive resource offers with varying levels of consolidation, meaning the GPUs offered may be located on a single machine, within a single rack, or spread across the network. Dally employs the well-established delay scheduling strategy to handle these offers. Under this scheme, a job has the option to decline offers that do not meet its consolidation preference as described in Algo. 1.  Initially, a job may reject all offers that do not place its GPUs on the same machine until a machine-level delay timer has elapsed while waiting (starving) for resources (line 10 in Algo. 1).  Once the machine-level delay timer elapses, the job  will commence accepting offers that position all its GPUs within a single rack  until the rack-level delay timer has also elapsed (line 18 in Algo. 1). Subsequently, at this point, the job will accept any resource offer without considering consolidation (line 21 in Algo. 1).
Note that a job that cannot fit into a single machine will have a machine-level delay timer set to 0, and similarly, a job that cannot fit into a single rack will have both delay timers set to 0.\par 

The underlying principle is to delay the placement of jobs in anticipation of better consolidation options becoming available in the future.  This delay-based approach strikes a balance between the high queuing times associated with consolidation and the high communication latency of non-consolidation. In traditional schedulers, these delay timers are typically configured by system administrators and often left at their default values of a few seconds. Based on our empirical observations, we establish a default value of 12 hours for machine-level consolidation and another 12 hours (or 24 hours in total) for rack-level consolidation. Given the extended duration of DL jobs (days and weeks), it is reasonable to consider delay timers ranging from half a day to a full day, in contrast to the more typical timer values of just a few seconds used in Spark~\cite{spark} and YARN~\cite{yarn} for shorter (few minutes) jobs. Generally, system administrators adjust timers based on historical resource contention data of the cluster. But, to eliminate the need for manual timer adjustments, we introduce an ``auto-tuner'' that automatically optimizes these delay timers. 
Remember that this is an MLaaS provider and will often encounter ``new'' models.

\begin{algorithm}
\caption{2 Get Tuned Timers}
\label{algo: get_tuned_timers}
\begin{algorithmic}[1]
\Require GPU demand $G_d$
\Ensure Tuned delay timers

\State $mc\_starvation\_times$ = $\mbox{get\_delay\_list}(machine, G_d)$

\For{$\text{time}$ in $\text{mc\_starvation\_times}$}
    \If{$\text{time} > \text{HISTORY\_TIME\_LIMIT}$}
        \State remove $\text{time}$ from $\text{mc\_starvation\_times}$
        \EndIf
\EndFor
    
\State $rack\_starvation\_times$ = $\mbox{get\_delay\_list}(rack, G_d)$

\For{$\text{time}$ in $\text{rack\_starvation\_times}$}
    \If{$\text{time} > \text{HISTORY\_TIME\_LIMIT}$}
        \State remove $\text{time}$ from $\text{rack\_starvation\_times}$
        \EndIf
\EndFor


\State \Return $\text{average}(\text{mc\_starvation\_times}) + 2 \times \text{standard\_deviation}(\text{mc\_starvation\_times}),$

\State \hspace{1.8cm} $\text{average}(\text{rack\_starvation\_times}) + 2 \times \text{standard\_deviation}(\text{rack\_starvation\_times})$
\end{algorithmic}
\end{algorithm}

\subsubsection{Auto-tuner}
We introduce an auto-tuner module in Dally aimed at automatically {\em fine-tuning} delay timers. This tuner relies on ``moving averages''  derived from historical job waiting (starvation) times associated with specific consolidation and GPU demands. The core concept revolves around leveraging historical data to determine an optimal waiting (delay) duration for securing a favorable job consolidation. When a job accepts a resource request, it updates a ``global list'' pertaining to the consolidation level (machine or rack) and the job's GPU demand, with the time it waited for before accepting a resource offer (lines 7 and 15 in Algo. 1). Dally maintains  comprehensive lists, tracking job waiting times for each combination of consolidation and GPU demand. In a DL  cluster, the GPU demands usually align with multiples of 2, typically ranging from 5 to 10 types. As a result, this approach leads to a manageable 2-20 lists for machine-level and rack-level consolidation paired with GPU demands.\par

\begin{figure}
    \centering
    \includegraphics[width=1\linewidth]{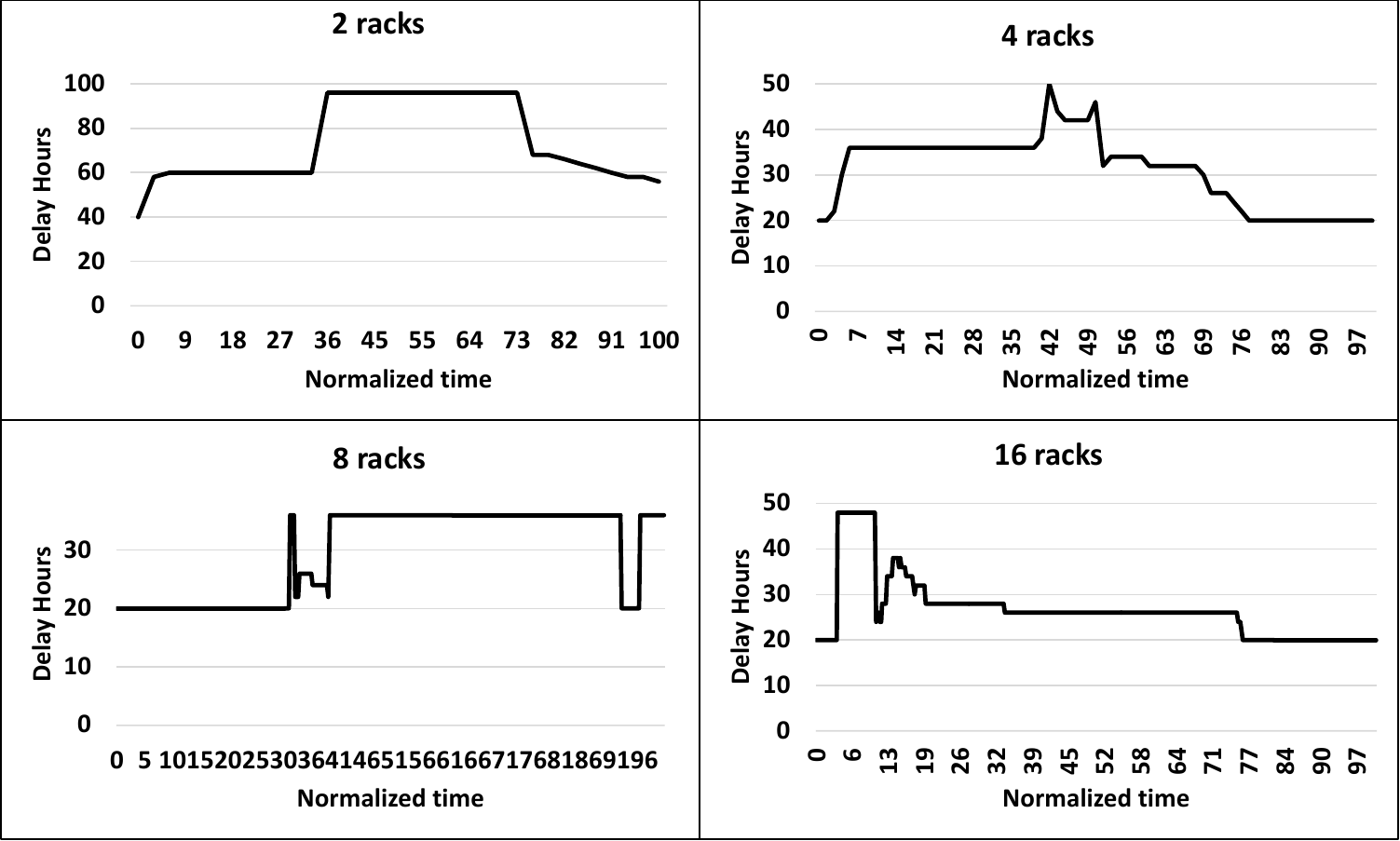}
    \caption{Auto-tuning timeline for rack-level delay timers.}
    \label{fig:auto-tune-timeline}
\end{figure}

To calculate the moving averages (described in Algo. 2), Dally mandates users to define the size of the wait-time list, denoted as the ``history limit'' 
(sliding window size) for the auto-tuner. The maintenance of the wait-time lists involves discarding values (utilized in average calculation) that exceed the specified history time limit (lines 4 and 10 in Algo. 2). Whenever a job receives a resource offer, the delay timer is calculated by averaging the wait-time list corresponding to the consolidation level (machine or rack) and GPU demand of the job, and then adding two sample standard deviations\footnote{Two standard deviations above mean corresponds to 95\% confidence and is commonly made in network system performance evaluation \cite{2-sigma}.} (similarly estimated using the same wait-time list). 
This value is then used in determining the optimal duration for a job to wait for a favorable consolidation (lines 13 and 14 in Algo. 2). 
We present a timeline depicting the adjustment of delay timers (rack level) for a typical P95 DDL job in Fig.~\ref{fig:auto-tune-timeline}. 
In the figure, the auto-tuner modifies delay timers by increasing or decreasing their values based on resource contention. Clusters with higher resource  contention (smaller clusters) have larger delay values.
Over time, the auto-tuner effectively ``learns'' the optimal duration to wait as demand delays are updated.

Note that it is essential to tailor the history limit according to the cluster size. Specifically, in larger clusters, a smaller history time limit is advised. This is simply because more jobs get placed over time in a larger cluster compared to a smaller cluster.

\subsection{Simulation Methodology}
In this work, we develop a high-fidelity DDL cluster simulator -- Artificial Intelligence System Simulator (\artist) -- capable of accurately replicating network slowdowns induced by specific job placements throughout a job's execution. 

\subsubsection{Simulator design}
\begin{figure}
    \centering
    \includegraphics[width=0.5\linewidth]{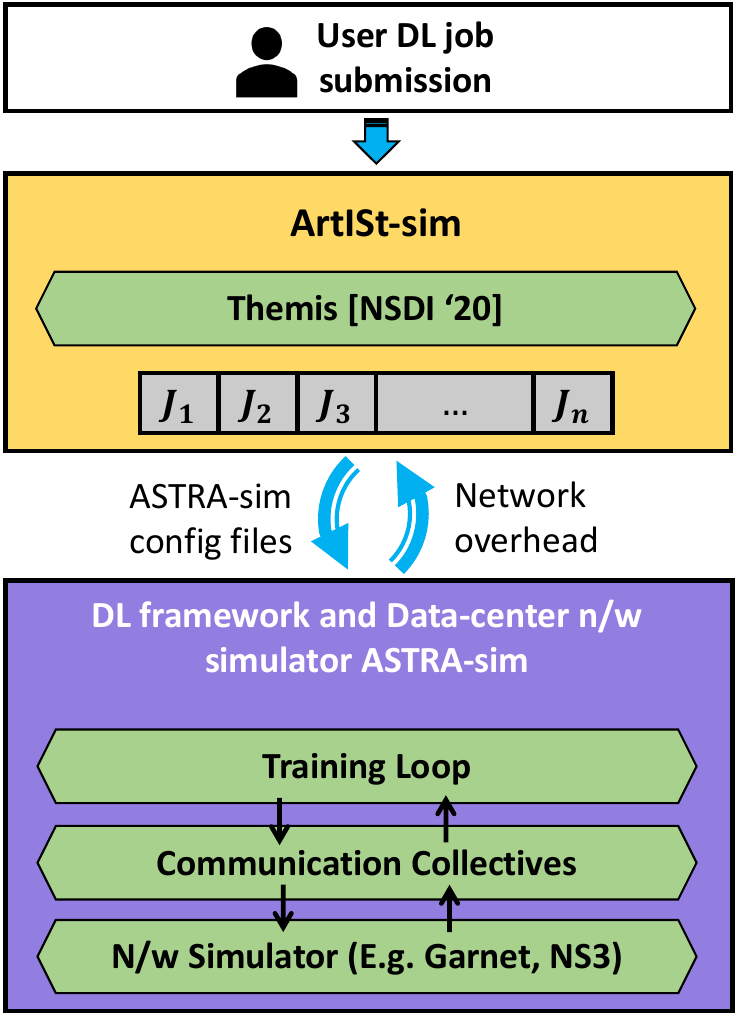}
    \caption{Simulation design. \label{fig:sim}}
\end{figure}

\artist builds upon a prior work, Themis~\cite{themis-nsdi}, which is a DL cluster scheduler simulator, and \astra. The Themis simulator simulates multiple jobs in a data center like a top level cluster scheduler such as Kubernetes~\cite{kubernetes}. The choice of Themis is based on its unique capability to simulate DL jobs at an iteration level -- a feature not present in other simulators like~\cite{lucid-asplos23, chronus-socc21, tiresias}.  This granular simulation approach enables us to precisely track ``job progression''  for preemption and subsequent placement decisions throughout the job's lifetime. However, as previously discussed, the network slowdown in the original work is encoded statically i.e., hard coded static values for rack and inter-rack placements with no provision to accommodate any other type of inter node network connection. To address this limitation, in \artist, we modified Themis to invoke \astra for each new job placement which can now be of any given type (e.g., NVSwitch, RoCE, Infiniband, ethernet etc.) as described by the user (refer to Fig.~\ref{fig:sim}). \par

For executing \astra (a single DDL simulator) with specific network placements,  \artist dynamically generates the necessary input -- workload, system, and network topology files -- and passes them to \astra (downward blue arrow in Fig.~\ref{fig:sim}) to simulate a specific job.
These generated files delineate the ``portion'' of the datacenter network utilized by the worker GPUs. For instance, in Fig.~\ref{fig:nw_arch}, the grey-colored job occupies 8 GPUs across two machines connected via a network link. To simulate \astra for such a placement, \artist produces the 8-GPU configuration files with a two-tier hierarchy representing the placement.  The  configuration files specify that the intra-machine GPUs are connected via NVSwitch in a ring topology, and the two machines are linked via a network (Ethernet) switch using a switch topology. Other \astra configurations, such as collectives, are also dynamically generated.
\astra then accurately simulates the specific job for a single DDL iteration based on the placement and relays the communication latency back to \artist (upward blue arrow in Fig.~\ref{fig:sim}). This latency is then utilized to simulate the job's progression (iterations) until it is either interrupted due to preemption or successfully completes its execution.\par
\begin{figure}
    \centering
    \includegraphics[width=.8\linewidth]{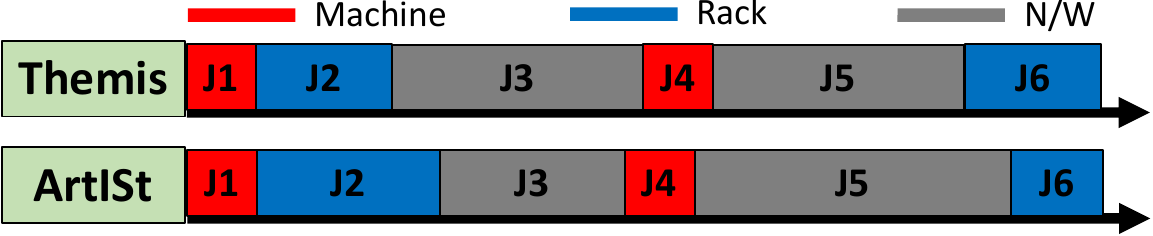}
    \caption{Job communication latency timeline. \label{fig:sim-timeline}}
\end{figure}

In Fig~\ref{fig:sim-timeline} we show the timeline of communication latency applied by Themis and \artist for 6 jobs. Themis simply applies a fixed latency penalty based on the ``network tier" (machine, rack or network) of the jobs placement. 
However, \artist dynamically calculates the latency of each job based on the specific placement at the given time for a high fidelity simulation. 
In the figure, \artist assigns a unique latency for each of the jobs.
Note that the compute time per iteration is extracted from the workload trace, and \astra serves solely to calculate the network latency for the specific placement.\par



\subsubsection{Simulation calibration and accuracy}
To ensure the fidelity of our simulations, we employ a calibration process that incorporates real-world measurements into \artist. We execute each model listed in Table~\ref{tab:overhead_table} on an 8-GPU machine connected via NVSwitch and calculate the associated network latency. Along with this, we simulate the model with the same network configurations (NVSwitch) in \astra. Subsequently, we scale the workload configuration file of \astra (for each model) to eliminate the discrepancy between simulated and actual network latency. For each DNN workload, the error in simulation is brought down to be within 1\%. These modified workload files are then used for all \astra calls. The ``calibrated'' simulator is now ready to simulate latency arising from rack and network-level placements. Also, for the verification of the various baseline scheduler implementations in \artist, we rely on the fact that Themis has already been validated in~\cite{themis-nsdi} with real world experiments i.e. each baseline in the simulator was also run in real world setup. Note that only a simulator version of the Tiresias source code is publicly available and no source code of Gandiva is publicly available. For \astra, the fidelity has been validated in various prior work such as~\cite{rashidi2020astra, won2023astrasim20, rashidi2022themis, Akella22}. 
Additionally, we faithfully reproduced the design principles of the Tiresias and Gandiva schedulers as outlined in their respective publications. 

\subsubsection{System Workloads}
We employ the cluster job traces made available by SenseTime~\cite{sensetime} through the work presented in~\cite{lucid-asplos23}. These traces are more recent than those utilized in~\cite{philly_trace_analysis} and \cite{weng2022mlaas}, and also provide a more realistic representation of a real-world DL cluster with multiple jobs being submitted over an extended period. This trace encompasses individual job arrivals alongside pertinent metadata information.

\subsubsection{Implementation}
We have implemented \artist in about 2000 lines of Java. \artist requires the network configuration parameters for intra-machine, intra-rack, and inter-rack interconnects. These parameters encompass the network topology, bandwidth, latency, collective communication algorithm, and other relevant details. 
Additionally, the user needs to specify the number of machines per rack and the total number of racks in the cluster.

\section{EXPERIMENTAL METHODOLOGY} \label{sections: methodology}
This section details the workloads, setup, baselines and metrics used in our evaluation.
\subsection{Workloads}
In our experiment, we employ real-world production traces obtained from SenseTime. 
We randomly select 500 DL/DDL jobs for batch and about 400 DL/DDL jobs for Poisson workloads from the trace,\footnote{From the original SenseTime trace, we replace EfficientNet with AlexNet
 for a high skew workload} which utilize the models specified in Table \ref{tab:overhead_table}.
The table encompasses a diverse range of network latency, providing ample coverage for testing a DDL cluster executing various network-sensitive DNN models. 
The cluster trace furnishes the compute requirements for individual jobs, including the time per iteration, the necessary number of iterations, and the quantity of GPUs demanded. 
We assume a data-parallel distributed training paradigm.
Note that both our batch and Poisson experiments push our system into a congested network  regime (a common occurrence in multi-tenant environments~\cite{philly_trace_analysis}), which is the scenario of interest.


\subsection{Cluster Setup}
In our experiments, we adopt the configuration of a three-tier network data center. 
Each machine is equipped with eight GPUs interconnected through NVIDIA NVSwitch. 
Within each rack, eight machines are connected via the NVIDIA Quantum switch, and the racks across the cluster are interlinked through the NVIDIA Spectrum network switch. 
To set up our network simulations in \astra, we rely on the QoS data advertised by NVIDIA \cite{nvidia-quantum, nvidia-spectrum}. 
Our experimentation involves manipulating the number of racks, specifically using 2, 4, 8, and 16 racks, to observe the impact of varying compute and network resources.



\begin{figure}
    \centering
    \includegraphics[width=1\linewidth]{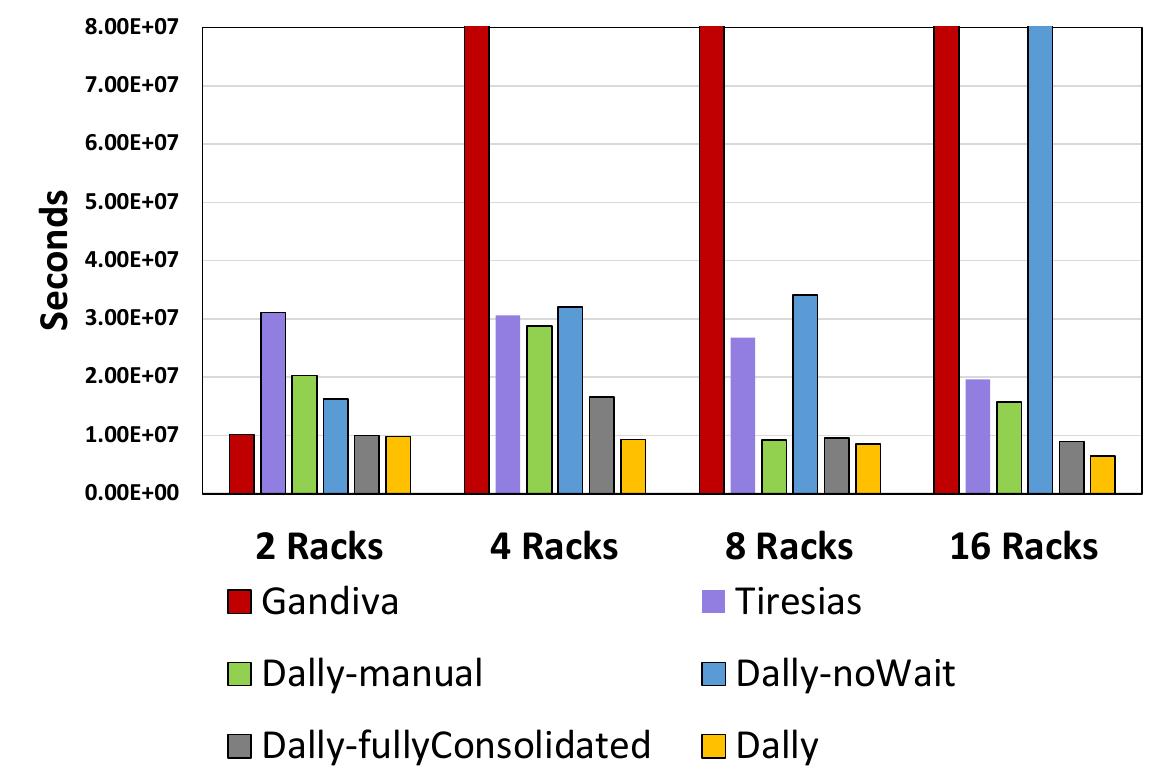}
    \caption{Batch arrival results -- Makespan}
    \label{fig:batch-results-makespan}
\end{figure}

\begin{figure*}
    \centering
    \includegraphics[width=1\linewidth]{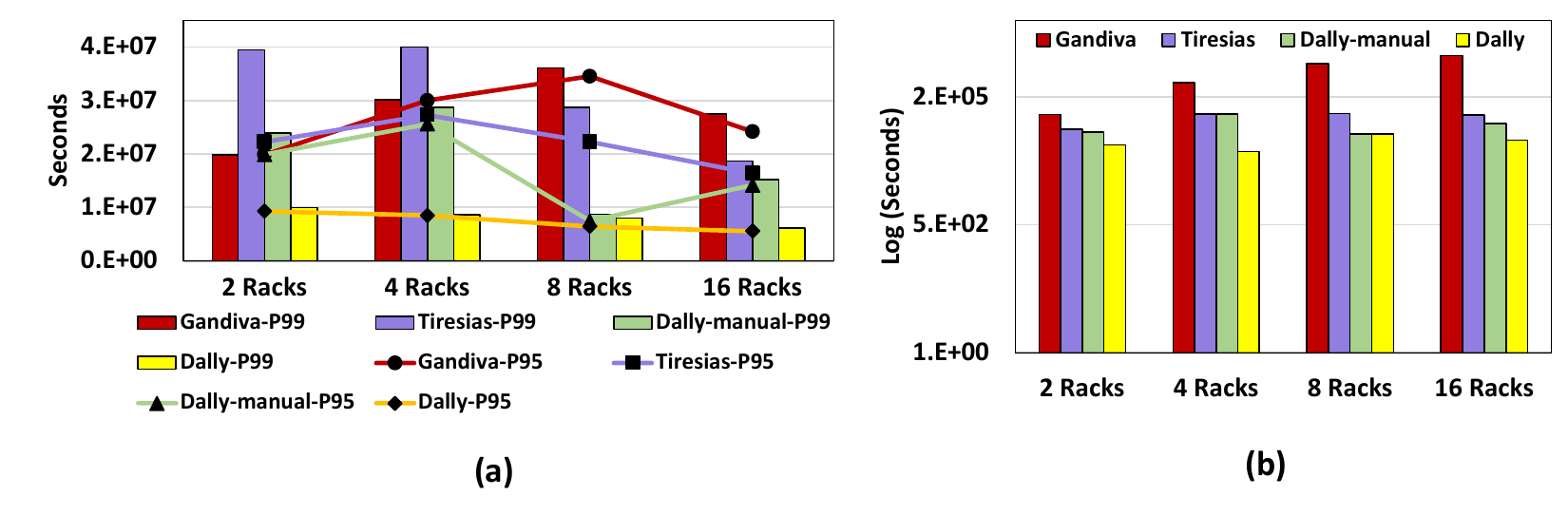}
    \caption{Batch arrival results -- (a) Tail queueing delay; (b) Average communication latency.}
    \label{fig:batch-results}
\end{figure*}

\subsection{Baselines}
In our evaluations, we performed experiments with the following schedulers: 

\begin{enumerate}
\item \emph{Gandiva}: Being network-agnostic, Gandiva, as one of the initial DL cluster schedulers, exhibits sub-optimal performance. It addresses this limitation by migrating jobs to more favorable GPU consolidation whenever resources become available.

\item \emph{Tiresias}: The state-of-the-art DL cluster scheduler currently employed adopts a strategy of consolidating job placement based on their concept of skew in DNN models. This scheduler possesses partial knowledge of the network, and the consolidation is stringent for skewed DNN models.

\item \emph{Dally-manual}: This represents a modification of our proposed scheme where delay timers for delay scheduling are manually configured. This aligns with the approach utilized by contemporary state-of-the-art big-data cluster schedulers, such as YARN.  

\item \emph{Dally-noWait}: Yet another modification of our scheme where the delay timers are set to 0 i.e., the scheduler does not wait for a consolidated placement if none is available.

\item \emph{Dally-fullyConsolidated}: This final modification of our scheme disregards delay timers and waits as long as needed for the most consolidated placement.

\end{enumerate}

Note that the preemption mechanism remains consistent with our proposed scheme Dally with auto-tuning. We use this baseline to demonstrate the effectiveness of our delay timer auto-tuner.

\subsection{Performance Metrics}
We define the following metrics to measure performance benefits of a job.

\begin{enumerate}
\item \emph{Makespan}: The time elapsed between arrival of the first job to the completion of the final job in the cluster. This is also a proxy for cloud-spend as public cloud bills its tenant based on wall-clock time or makespan occupying its resources. Hence, the monetary expenditure is always a multiple of it.

\item \emph {JCT}: The job completion time (JCT) is the elapsed time between job arrival and completion in the cluster.

\item \emph {Queueing delay}: Time spent by a job in the wait queue.

\item \emph {Communication latency}: Exposed communication time.

\end{enumerate}

\section{EVALUATION} \label{sections: evaluation}
In this section, we assess Dally through extensive simulations driven by large-scale traces, following the outlined design. 
It is important to emphasize that our evaluations leverage the most recent generation of network hardware, necessitating the use of simulations as the exclusive option.

\begin{figure*}[htbp]
    \centering
    \subfloat[2 racks]{{\includegraphics[width=0.24\linewidth]{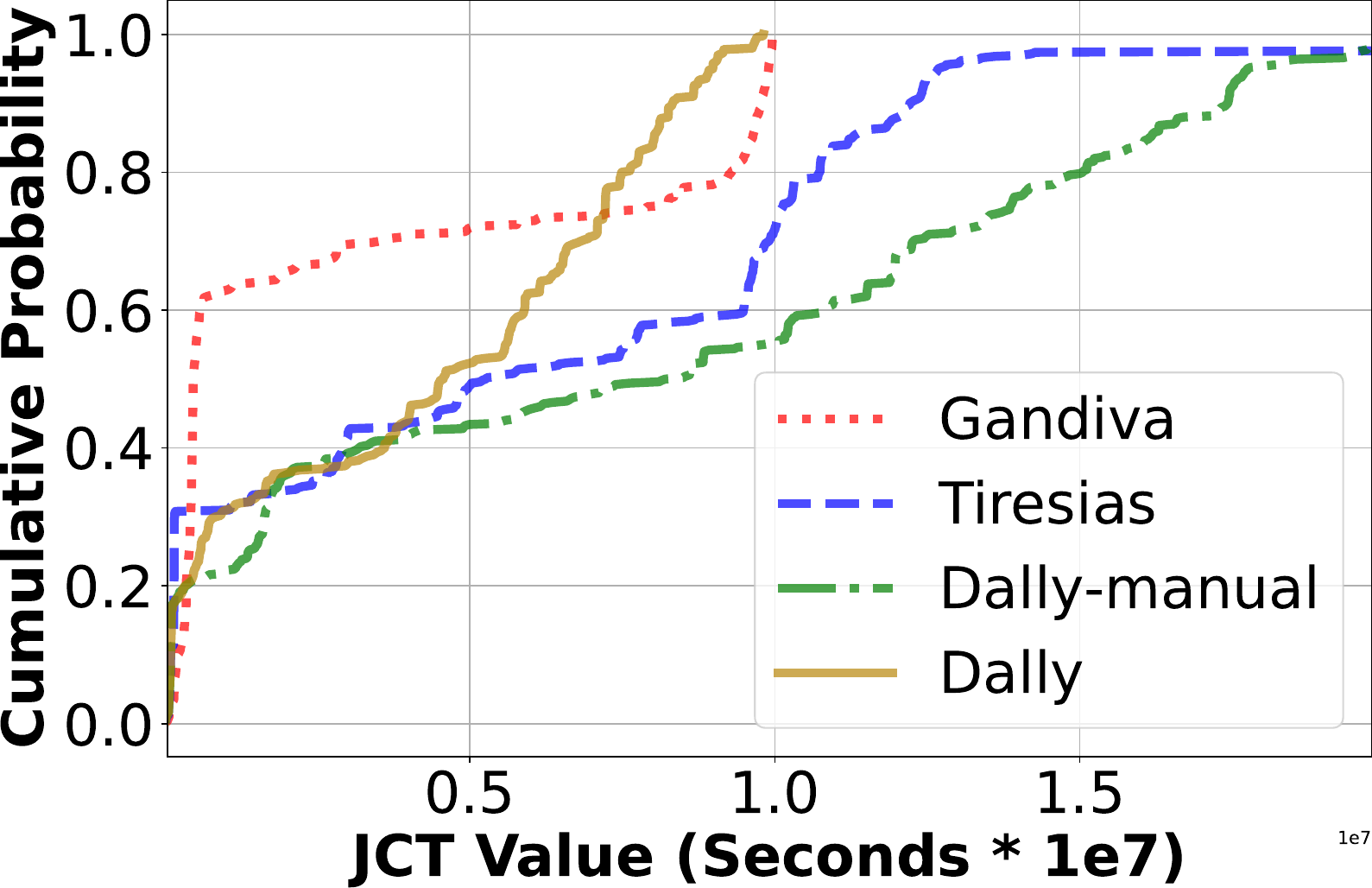} }}
    \subfloat[4 racks]{{\includegraphics[width=0.24\linewidth]{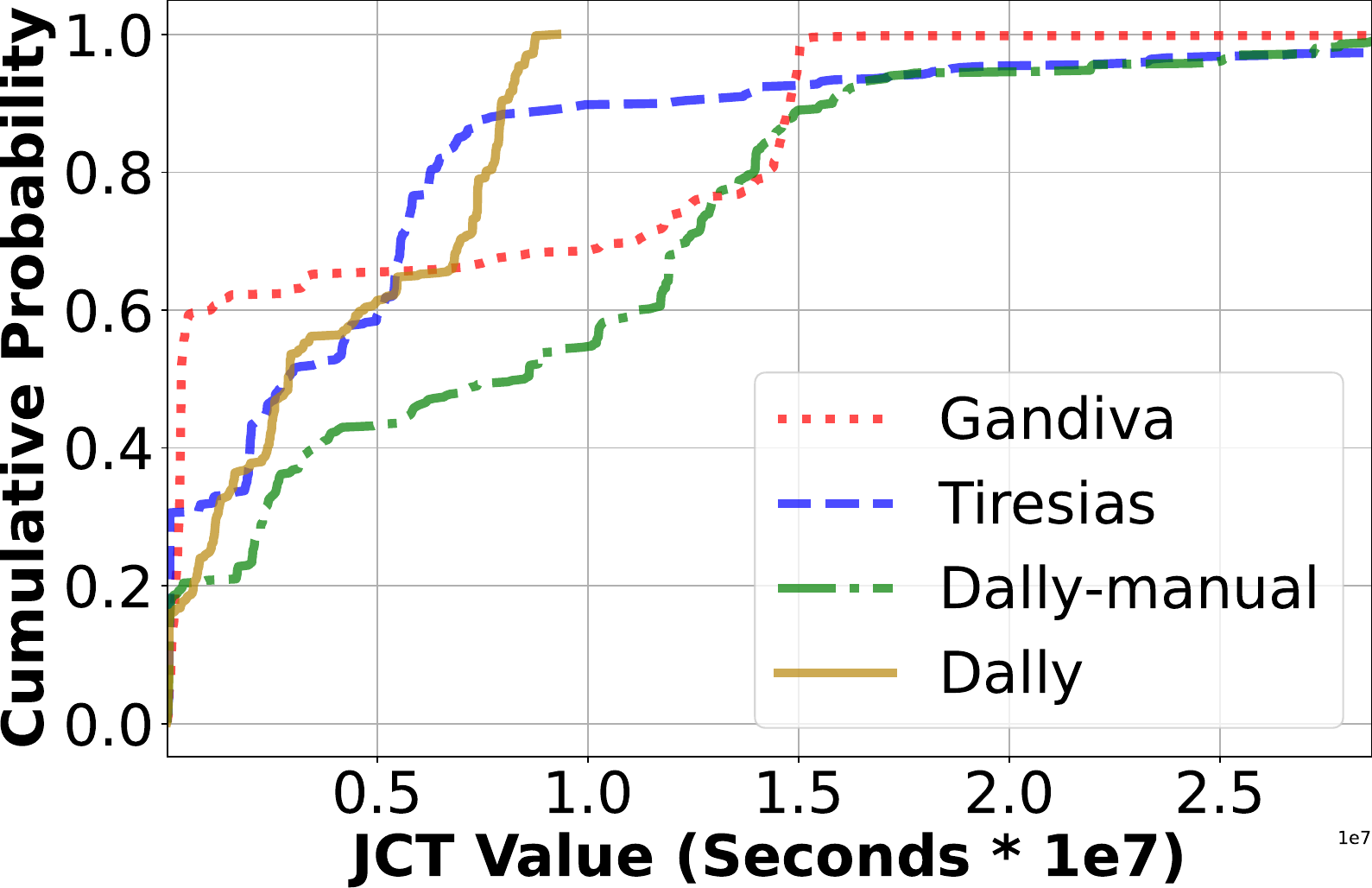} }}
    \subfloat[8 racks]{{\includegraphics[width=0.24\linewidth]{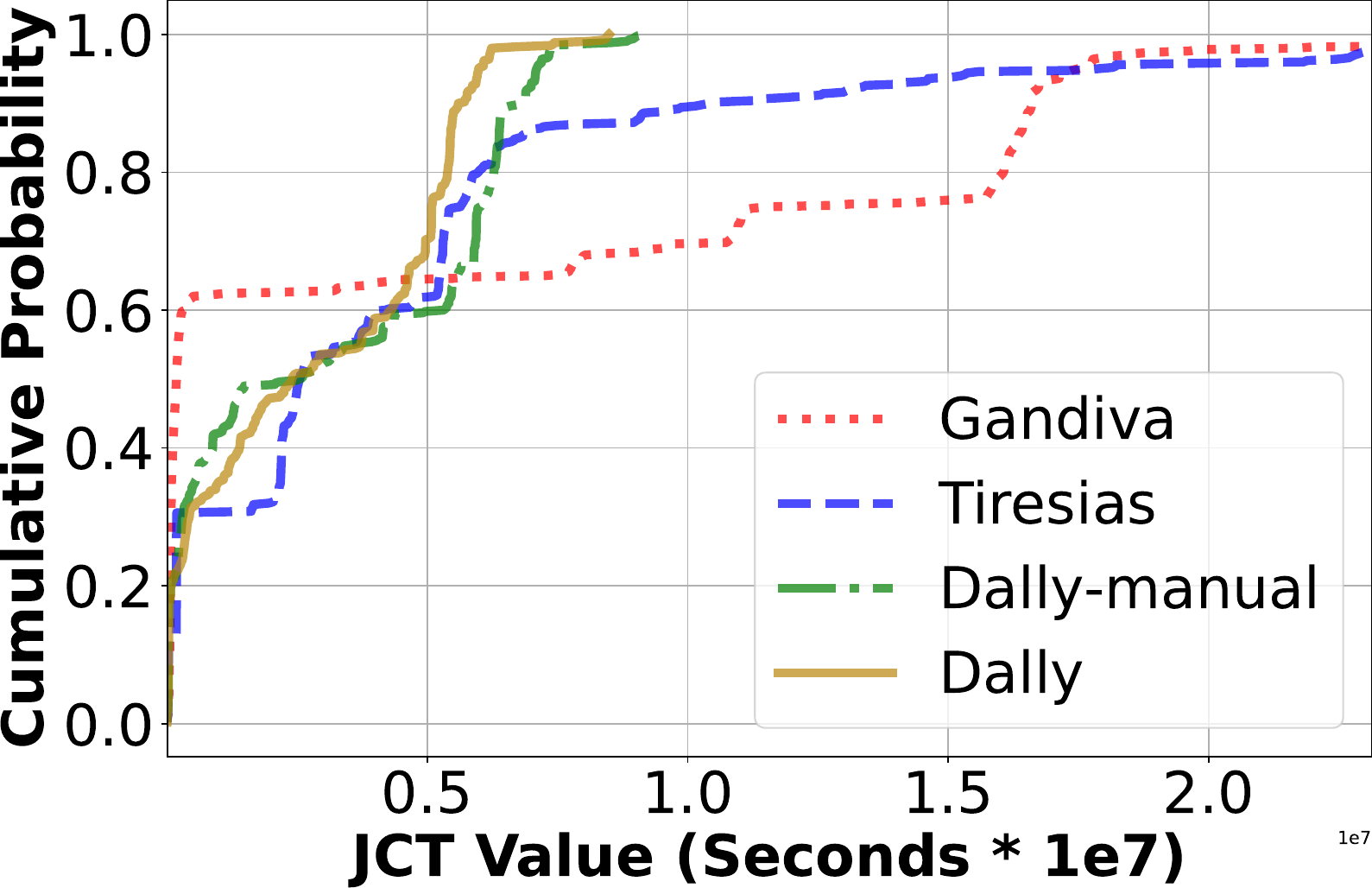} }}
    \subfloat[16 racks]{{\includegraphics[width=0.24\linewidth]{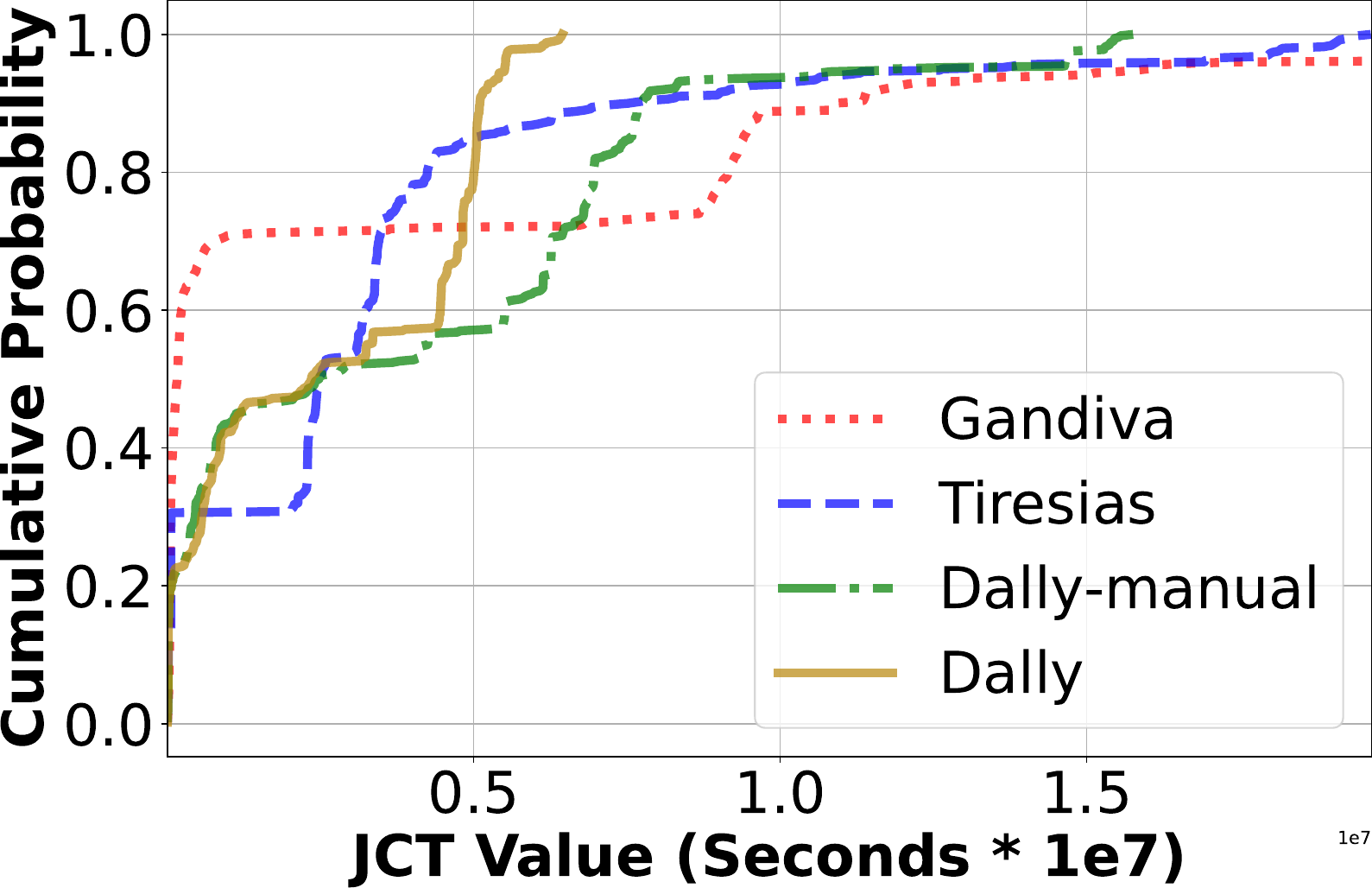} }}
    \caption{JCT CDF -- Batch arrival.}
    \label{fig:jct-cdf-batch}
\end{figure*}

\begin{figure*}[htbp]
    \centering
    \subfloat[2 racks]{{\includegraphics[width=0.24\linewidth]{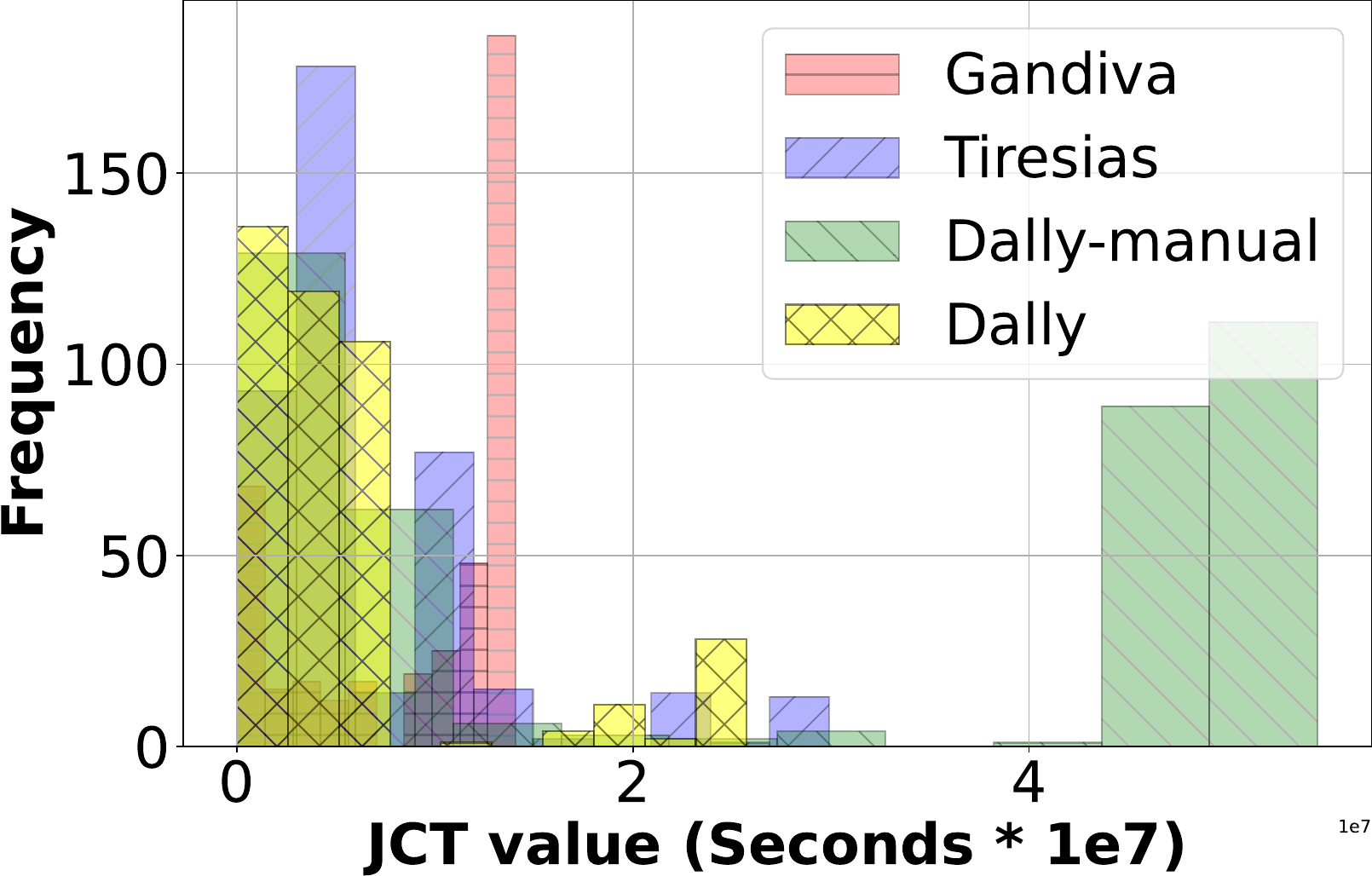} }}
    \subfloat[4 racks]{{\includegraphics[width=0.24\linewidth]{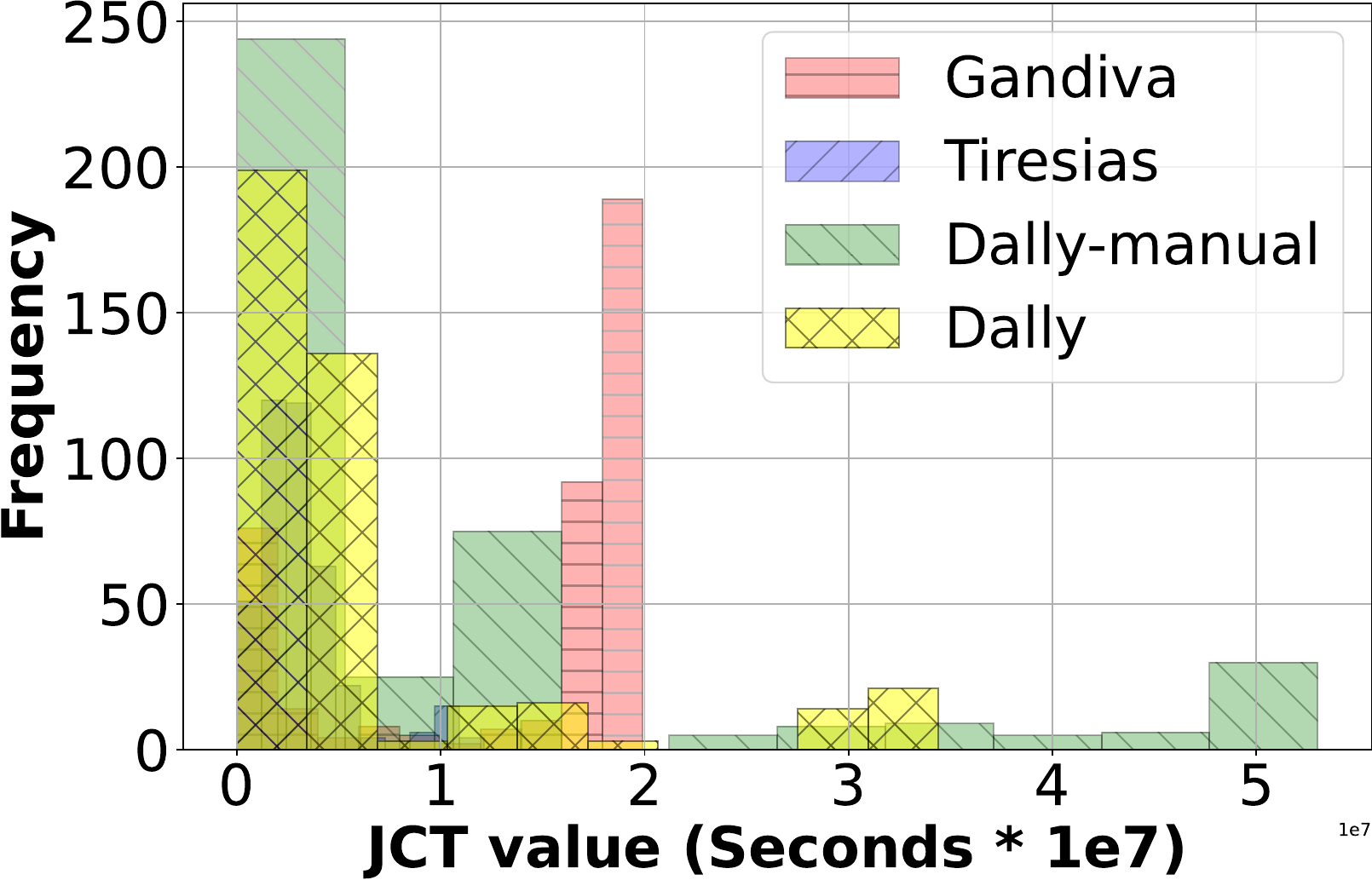} }}
    \subfloat[8 racks]{{\includegraphics[width=0.24\linewidth]{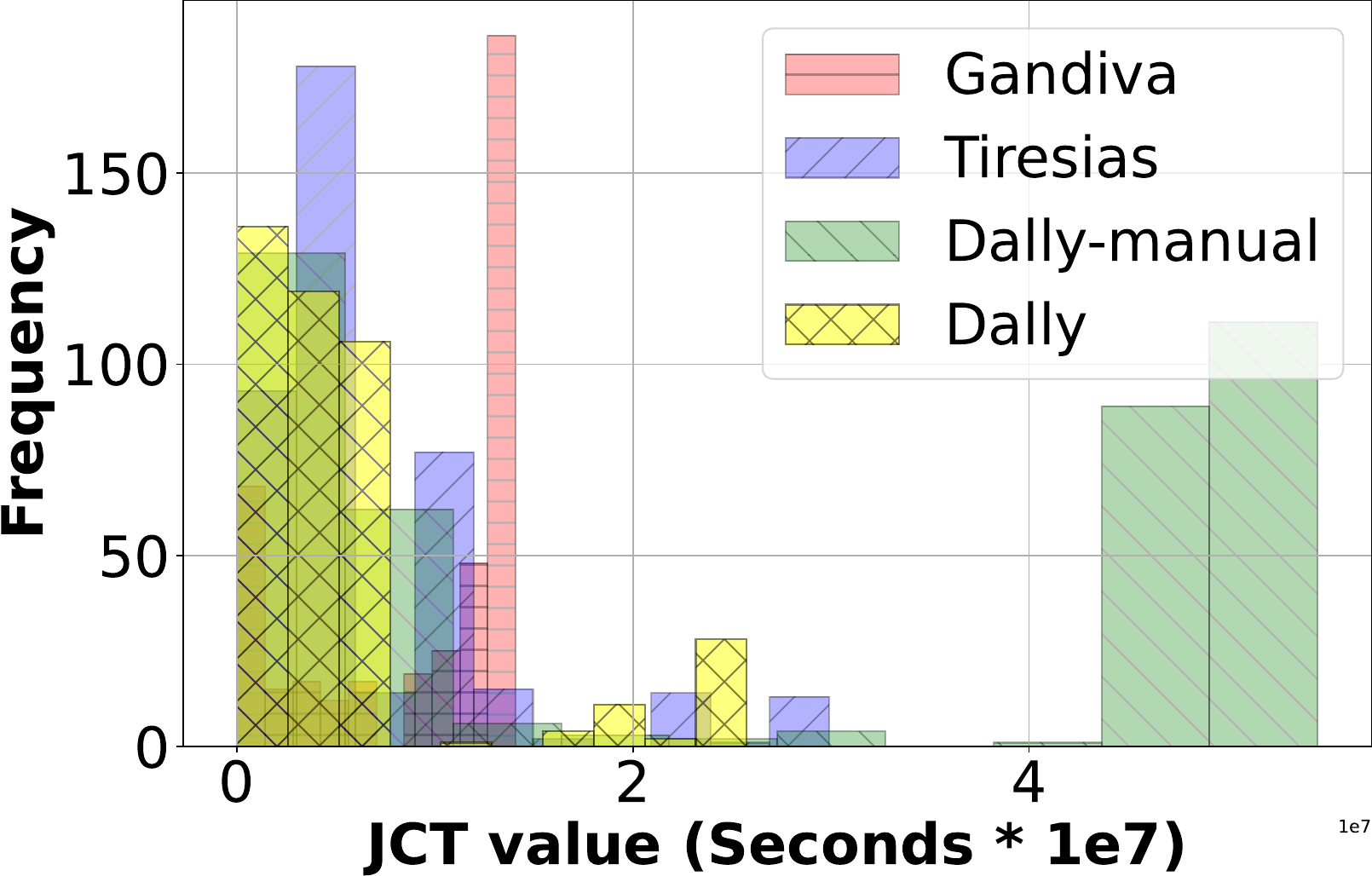} }}
    \subfloat[16 racks]{{\includegraphics[width=0.24\linewidth]{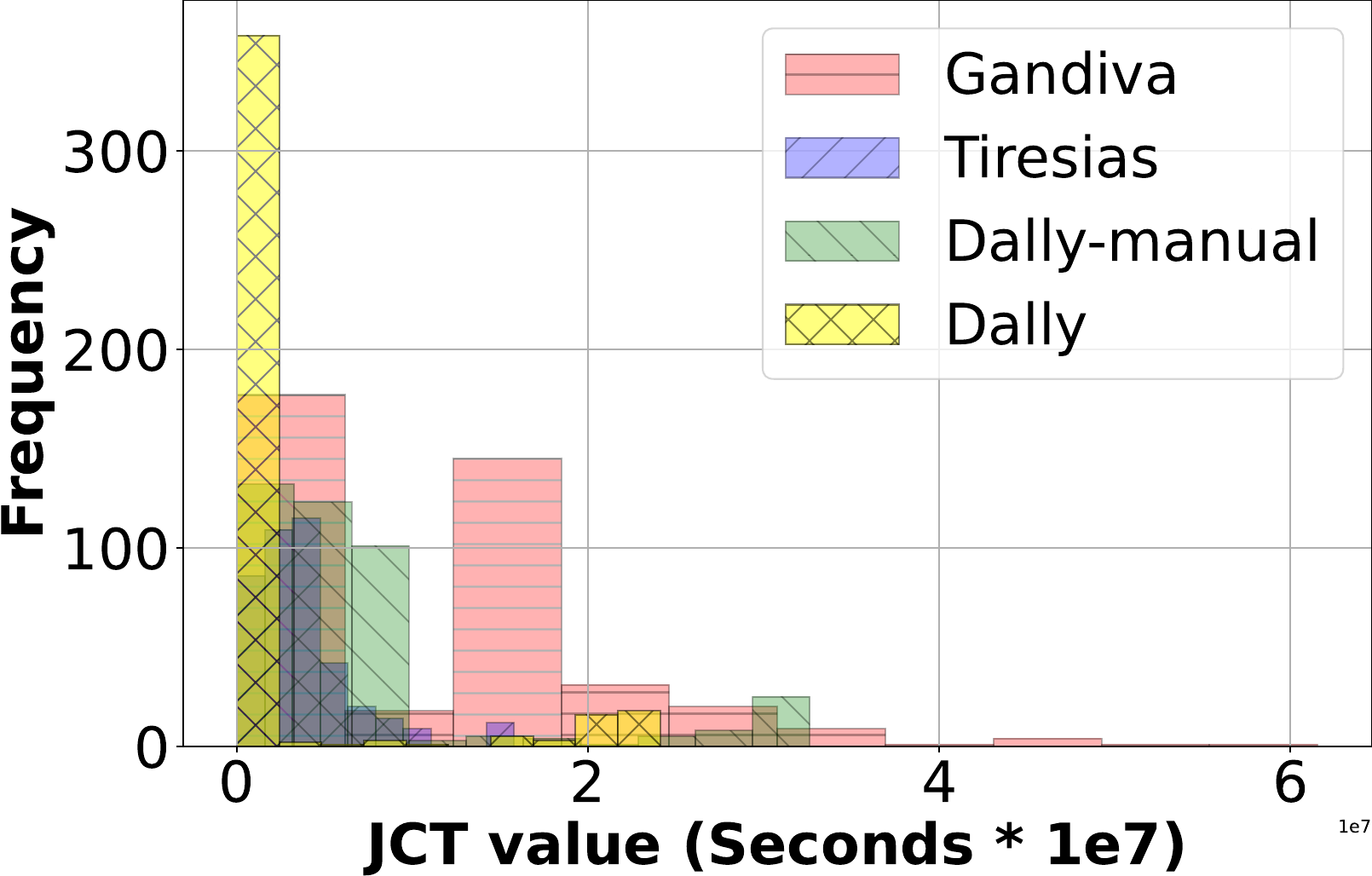} }}
    \caption{JCT histogram (P95) -- Poisson arrival.}
    \label{fig:jct_poisson_hist}
\end{figure*}

\subsubsection{Makespan}
Fig.~\ref{fig:batch-results-makespan} illustrates the improvement in makespan. Notably, Dally exhibits an improvement of up to 69\% compared to Tiresias and up to 92\%, compared to Gandiva. This improvement is attributed to the implementation of delay scheduling, effectively relaxing stringent consolidation, thereby decreasing queueing delay and optimizing JCT by judiciously awaiting resource offers with favorable consolidation. A loosening  essentially implies a job being positioned in less favorable network tiers, a scenario which may not be feasible in Tiresias. The manual configuration of delay timers in Dally-manual results in sub-optimal waiting, as it cannot dynamically {\em tune} its delay timers based on contention.  The static nature of the delay time fails  to consider GPU demand and resource contention in the cluster, leading to inferior performance compared to Dally, while still outperforming Tiresias and Gandiva due to reduced queuing delays. Dally-noWait experiences high network latency, leading to poorer performance compared to Tiresias. However, in the case of two racks with limited inter-rack communication, it outperforms Tiresias in terms of makespan, thanks to our novel network-sensitive preemption policy. Dally-fullyConsolidated performs worse than Dally but still surpasses Tiresias, as it does not utilize the delay scheduling policy. These results clearly highlight the effectiveness of both our preemption and delay scheduling strategies.



\subsubsection{Queuing Delay}
The primary contributor to the significant improvement in makespan is the reduction in queueing delays experienced within the cluster.  Although Dally reports an average queueing delay improvement ranging from 18\% to 31\%, depending on the cluster size, the cluster's makespan is more notably influenced by the \>P95 and \>P99 JCTs (i.e. the 95th and 99th complementary percentiles of JCT distribution).  Dally demonstrates a remarkable improvement of 58\% -- 71\% in \>P95  and 67\% -- 78\% in \>P99 queuing delay compared to Tiresias (Fig.~\ref{fig:batch-results} (a)), which results in the substantial improvement in makespan. The reduction in queuing delay can be attributed to the less stringent consolidation policy of Dally compared to Tiresias, although not as relaxed as Gandiva's policy. Yet, Dally  improves upon Gandiva in tail (\>P95 and \>P99) queueing delay by as much as 81\%. In fact, Gandiva's network-agnostic scheduling policy leads to very high communication latency.  


\subsubsection{Communication latency}
\begin{figure}
    \centering
    \subfloat[Batch]{{\includegraphics[width=0.5\linewidth]{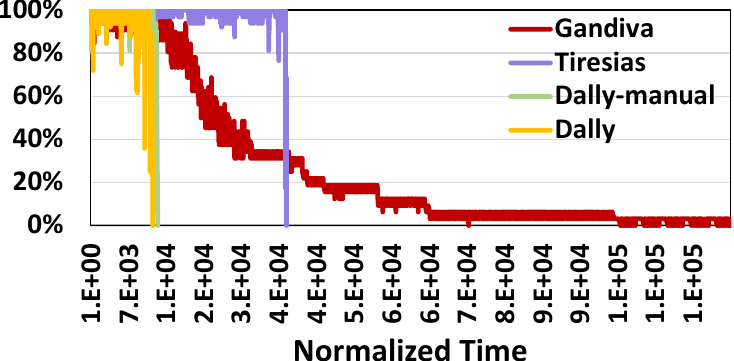} }}
    \subfloat[Poisson]{{\includegraphics[width=0.5\linewidth]{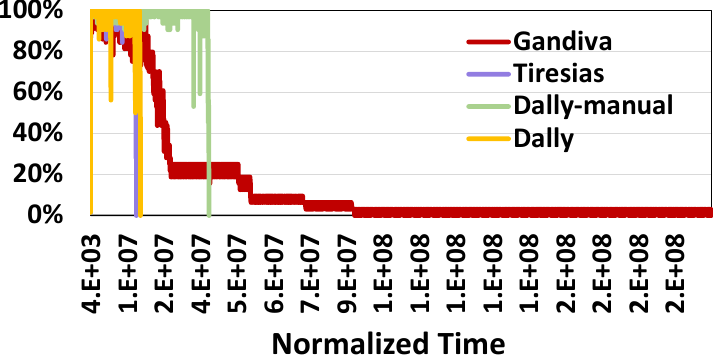} }}
    \caption{Cluster GPU utilization (8 racks).}
    \label{fig:jobs_cluster_util}
\end{figure}

Dally demonstrates the lowest communication latency, surpassing the nearest network-aware baseline, Tiresias, by a notable margin ranging from 53\% to 83\% (Fig.~\ref{fig:batch-results} (b)). This is due to Dally's preemption policy, which prioritizes providing better-consolidated placements to jobs suffering from sub-optimal placements or network sensitivity, thereby mitigating communication latency. Gandiva exhibits a slight performance difference in makespan compared to Dally in the initial 2-rack experiment, but experiences a substantial degradation as the number of racks increases.
This discrepancy arises because the 2-rack experiment involves fewer network connections and consequently fewer network placements. However, with an increase in the number of racks, the cluster experiences a rise in network connections and leads to jobs being placed on the network, resulting in elevated communication latency, as depicted in Fig.~\ref{fig:batch-results}(b). 
Additionally, this phenomenon results in a long tail in the number of job completions within the cluster, subsequently leading to low cluster utilization (Fig.~\ref{fig:jobs_cluster_util}).

\begin{figure}
    \centering
    \subfloat[Batch]{{\includegraphics[width=0.5\linewidth]{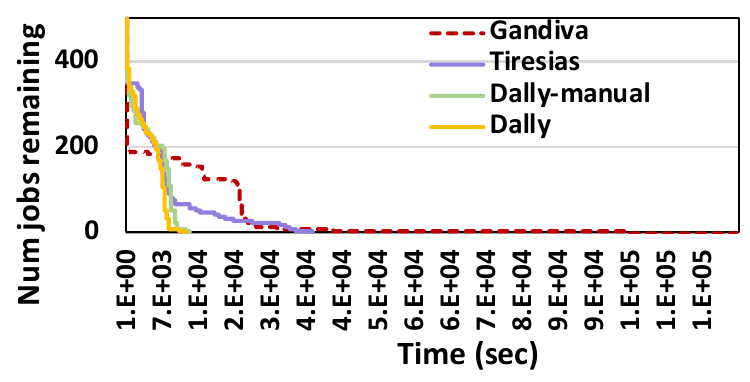} }}
    \subfloat[Poisson]{{\includegraphics[width=0.5\linewidth]{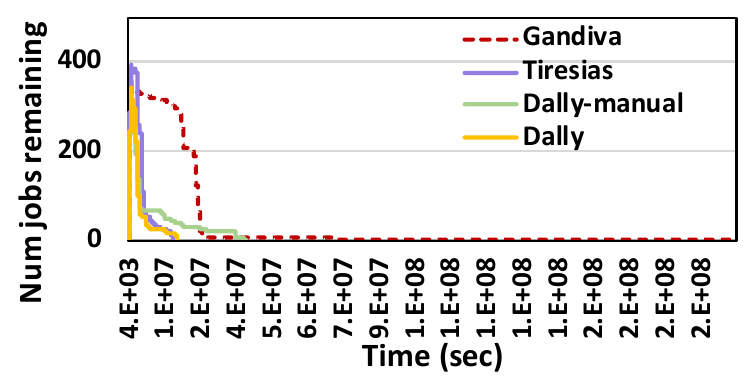} }}
    \caption{Number of jobs remaining (8 racks).}
    \label{fig:num_jobs_remain}
\end{figure}

\subsubsection{Cluster Utilization}
We explain this section through results from the 8 racks experiment but we get similar results for the other experiments as well. Fig.~\ref{fig:num_jobs_remain}  illustrates the number of remaining jobs in the cluster, highlighting Gandiva's notable tailing effect. Gandiva  jobs (including the \>P95 and \>P99) experience substantial  communication latency, resulting in diminished cluster utilization, as demonstrated in Fig.~\ref{fig:jobs_cluster_util}. Dally's efficacy in minimizing communication latency and queuing delays contributes to the highest job completion rates among all tested schemes, as shown in Fig.~\ref{fig:num_jobs_remain} (a proxy for cluster utilization), along with the lowest JCTs. Note, in real-world scenarios, users can leverage autoscalers \cite{aws-autoscaling} and release unused GPUs (below 100\% demand) back to the cloud for cost savings. For simplicity, this work excludes the complexities of autoscaling.

\subsubsection{Job Completion Time (JCT)}
\begin{figure}
    \centering
    \subfloat[Batch arrival]{{\includegraphics[width=0.5\linewidth]{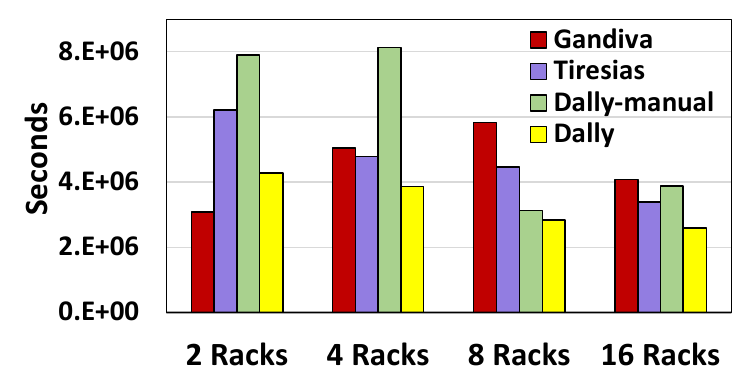} }}
    \subfloat[Poisson arrival]{{\includegraphics[width=0.5\linewidth]{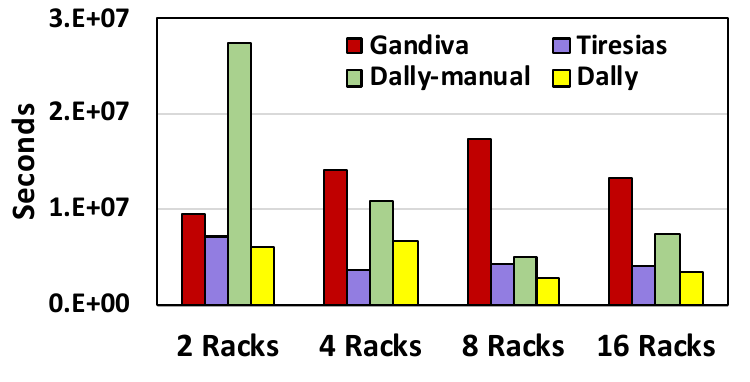} }}
    \caption{Average JCT.}
    \label{fig:jct_avg}
\end{figure}

To assess the JCT, we utilize traces with {\em both}  batch and Poisson job arrivals, recognizing that the average JCT can vary significantly based on the job arrival pattern. In Fig.~\ref{fig:jct_avg}(a), we present  the average  JCTs with batch arrivals, revealing that Dally achieves an improvement in average JCT ranging from 19\% to 36\% compared to Tiresias and 23\% to 51\% compared to Gandiva. Examination of the JCT Cumulative Distribution Functions (CDFs) in Fig.~\ref{fig:jct-cdf-batch} indicates that Dally reduces JCT for a majority of the job sizes. Notably, in the 2 racks experiment, Dally experiences a 39\% degradation against Gandiva, which performs better with fewer aggregate network placements. 
However, as the number of racks (and network connections) increases, Gandiva's performance degrades rapidly due to increased communication latency. Additional statistics for the JCT in the 8-rack experiment are presented in Table~\ref{tab:jct-8racks-batch-stats}, showcasing Dally's enhancements in tail JCT.\par

To assess Dally under Poisson arrival conditions, we conduct experiments analogous to those performed with batch arrivals, but with jobs exhibiting typical Poisson arrival times. 
Dally demonstrates improvements over Tiresias ranging from 16\% to 34\% and over Gandiva by 23\% to 51\% as shown in Fig.~\ref{fig:jct_avg}(b). However, in the 4-rack scenario, Dally experiences degradation compared to Tiresias due to the random arrival pattern inherent in a Poisson process.  A drawback worth noting is that  Tiresias exhibits superior performance in identifying optimal placements for this particular arrival pattern and network size (racks), improving upon its own average JCT in the 8-rack and 16-rack experiments. This indicates that the specific arrival pattern was well suited for Tiresias in the 4-racks case. Also, recall from Sec.~\ref{sections: preemption} that {\em Dally does not emphasize JCT when prioritizing jobs for GPU consolidation}. However, Fig.~\ref{fig:jct_poisson_hist}  presents histograms of JCTs (P95) with Poisson arrival times, illustrating Dally's improvement in the JCTs of the majority of the jobs. Detailed JCT statistics for the 8-rack Poisson experiment are provided in Table~\ref{tab:jct-8racks-poisson-stats}. We find Dally outperforming Tiresias in {\em both} average and median values by 35\% and 38\%, respectively, while being within 10\% of the tail. Dally also outperforms Gandiva in average and median JCT by 83\% and  88\% respectively and tail JCT by 76-92\%.
Note that our Poisson experiments represent peak usage scenarios to evaluate our scheme under network congestion.


\begin{table}[]
\small
\begin{tabular}{crrrr}
\hline
\multicolumn{1}{l}{} & \multicolumn{1}{l}{Gandiva} & \multicolumn{1}{l}{Tiresias} & \multicolumn{1}{l}{Dally-manual} & \multicolumn{1}{l}{Dally} \\ \hline
Average              & 5828719                     & 4461998                      & 3128566                          & 2828955                   \\
Median               & 183717                      & 2547080                      & 2552189                          & 2401390                   \\
P95                  & 17485823                    & 17841606                     & 7097646                          & 5983042                   \\
P99                  & 39214710                    & 25696135                     & 8868401                          & 8395214                   \\ \hline
\end{tabular}
\caption{JCT batch statistics in seconds (8 racks).}
\label{tab:jct-8racks-batch-stats}
\end{table}

\begin{table}[]
\small
\begin{tabular}{crrrr}
\hline
\multicolumn{1}{l}{} & \multicolumn{1}{l}{Gandiva} & \multicolumn{1}{l}{Tiresias} & \multicolumn{1}{l}{Dally-manual} & \multicolumn{1}{l}{Dally} \\ \hline
Average              & 17371514                    & 4329941                      & 4960727                          & 2831880                   \\
Median               & 22722917                    & 4212795                      & 2787855                          & 2595529                   \\
P95                  & 71362014                    & 15059080                     & 39298828                         & 16469069                  \\
P99                  & 208939883                   & 15390860                     & 39817762                         & 16681241                  \\ \hline
\end{tabular}
\caption{JCT Poisson statistics in seconds (8 racks).}
\label{tab:jct-8racks-poisson-stats}
\end{table}



\section{RELATED WORK} \label{sections: related-work} 

\subsection{DL Cluster Scheduling} 
A plethora of DL cluster schedulers have been developed recently with varying objectives such as JCT performance, fairness, and SLO compliance.
They utilize strategies like consolidation, migration, packing, resource reallocation, etc.
In terms of performance, network agnostic approaches like \cite{lucid-asplos23, ant-man} focus on packing multiple DNN models onto a single GPU, boosting parallelism and decreasing overall execution time. Our work, centered around network sensitivity to enhance performance, is orthogonal to such efforts, as our proposed techniques can be readily applied to GPUs simultaneously running multiple models. 
Further, works such as~\cite{pollux, optimus, dl2, smd, aonline} improve performance through resource reallocation. However, due to the high sensitivity of DDL to network latency, these schedulers can quickly degrade performance with an unfavourable GPU placement.

\cite{horus, muri, Co-scheML, harmony} are interference-aware schedulers and for fairness, \cite{themis-nsdi, gandiva-fair, ASTRAEA} have been proposed. \cite{chronus, ElasticFlow} are SLO-compliant schedulers and \cite{sc21ones, sc23easyscale} improve resource utilization through dynamically adjusting batch size and GPU workers respectively. Recently, Sia \cite{sia-sosp23} was introduced to reduce JCT. It employs integer linear programming (ILP) to determine the optimal GPU allocation for data-parallel deep learning jobs in a two-tier network. However, Sia overlooks the heterogeneity of network links and only accounts for two types of connections -- intra-machine and inter-machine. Additionally, solving an ILP for an n-tier network could be computationally intensive.
Further, to lower public cloud expenses, techniques such as \cite{cypress, paldia, faastloop} (for DNN) and \cite{sharma2021cash, kraken} (for data parallel workloads) have been proposed.

\subsection{Traditional Cluster Scheduling}
Early cluster scheduling was primarily designed for high-performance computing (HPC) workloads, leveraging the message passing interface (MPI)~\cite{open-mpi} for communication among workers. However, with the rise of big data, data centers quickly shifted towards data-parallel workloads such as MapReduce~\cite{hadoop}, Spark~\cite{spark}, and others. This shift led to the development of numerous job schedulers, specifically tailored for data-parallel tasks (e.g.,  see~\cite{delay_scheduling, DRF, grandl2014multi, grandl2016altruistic, grandl2016graphene, bhattacharya2013hierarchical, mantri-osdi10,  ousterhout2017monotasks} and the references therein).
Following this, need for more general-purpose schedulers capable of handling CPU-bound workloads or serving as cluster-wide schedulers became evident. This resulted in significant advancements in generic cluster scheduling. The examples include, but not limited to,~\cite{Mesos, Borg, kubernetes, schwarzkopf2013omega, ousterhout2013sparrow, delgado2016eagle}. However, deep learning workloads present unique challenges compared to both data-parallel and general CPU-bound tasks, particularly due to their reliance on exclusive GPU resource scheduling, low tolerance to network latency, and being highly repetitive.

\subsection{Network Scheduling}
Network scheduling deals with optimizing the network flow or the traffic in the cluster by optimizing either the network stack or the communication algorithms.
They are orthogonal to job schedulers and are typically run after resources are allocated to a job.
The scheduling strategy is largely driven by the datacenter workload. Some consequential work for data parallel workloads include~\cite{coflow, dens, corral, baraat, hedera}. For deep learning, recent work include \cite{cao2024crux, rajasekaran2024cassini, mahajan2023better, li2023accelerating, lao2021atp}.
\section{CONCLUSIONS} \label{sections: summary}
Training of DDL jobs on GPU clusters is expensive in both  public and private cloud environments. A significant part of this time is due to the communication latency between a hierarchical organization of GPUs in the underlying datacenter. This communication latency and the resultant overall training time can be minimized by a prudent allocation of GPUs, where a distributed training job is consolidated on physically proximal GPUs in proportion to its network sensitivity.\par

In this context, we propose Dally, a cost-effective DL cluster scheduling scheme which employs delay scheduling for job placement and consolidation. Dally incorporates a novel network-sensitive preemption technique to prioritize jobs with high network sensitivity, ensuring they receive more consolidated job offers. Additionally, we propose and integrate an auto-tuner into Dally to optimize delay timers according to varying GPU contention/demand for delay scheduling. To assess the performance of Dally, we introduce \artist, a cost-effective DL simulation platform tailored for simulating large DL clusters. Through extensive trace-driven simulations, we demonstrate that Dally can reduce end-to-end makespan by 69\% compared to a consolidation based policy, average JCT by up to 36\%, and communication latency by up to 83\% under congested networking conditions.  
  




\bibliographystyle{IEEEtran}


\bibliography{references,cloud,scheduling,DDoS,dnn,sysml}



\end{document}